\documentclass[sigconf, numbers]{acmart}
\usepackage{tcolorbox}
\usepackage{booktabs}
\usepackage{multirow}
\usepackage{multicol}
\usepackage{array} 

\AtBeginDocument{%
  \providecommand\BibTeX{{%
    \normalfont B\kern-0.5em{\scshape i\kern-0.25em b}\kern-0.8em\TeX}}}

\begin{document}

\title{Understanding the Human-LLM Dynamic: A Literature Survey of LLM Use in Programming Tasks}

\author{Deborah Etsenake}
\email{detsenake@uwaterloo.ca}
\orcid{1234-5678-9012}
\affiliation{%
  \institution{University of Waterloo}
  \streetaddress{200 University Ave W}
  \city{Waterloo}
  \state{Ontario}
  \country{Canada}
  \postcode{N2L 3G1}
}

\author{Meiyappan Nagappan}
\email{mei.nagappan@uwaterloo.ca}
\affiliation{%
  \institution{University of Waterloo}
  \streetaddress{200 University Ave W}
  \city{Waterloo}
  \state{Ontario}
  \country{Canada}}

\renewcommand{\shortauthors}{Etsenake and Nagappan.}

\begin{abstract}
 Large Language Models (LLMs) are transforming programming practices, offering significant capabilities for code generation activities. While researchers have explored the potential of LLMs in various domains, this paper focuses on their use in programming tasks, drawing insights from user studies that assess the impact of LLMs on programming tasks. We first examined the user interaction behaviors with LLMs observed in these studies, from the types of requests made to task completion strategies. Additionally, our analysis reveals both benefits and weaknesses of LLMs showing mixed effects on the human and task. Lastly, we looked into what factors from the human, LLM or the interaction of both, affect the human's enhancement as well as the task performance. Our findings highlight the variability in human-LLM interactions due to the non-deterministic nature of both parties (humans and LLMs), underscoring the need for a deeper understanding of these interaction patterns. We conclude by providing some practical suggestions for researchers as well as programmers.
\end{abstract}

\begin{CCSXML}
<ccs2012>
<concept>
<concept_id>10002944.10011122.10002945</concept_id>
<concept_desc>General and reference~Surveys and overviews</concept_desc>
<concept_significance>500</concept_significance>
</concept>
</ccs2012>
\end{CCSXML}

\ccsdesc[500]{General and reference~Surveys and overviews}

\keywords{Large Language Models, LLMs, User studies, User evaluations, natural language processing, programming tasks, usability studies}

\maketitle

\section{Introduction}
In recent years, Large Language Models (LLMs) have surged in prominence. Trained across diverse fields like computing, literature, and psychology, LLMs offer versatility for various tasks, including natural language programming (code generation) and code analysis. This rise in LLM usage has spurred extensive research exploring, their efficiency across different tasks \cite{Arghavan2023JSS,Ozturk2023EICC,Byun2023ACM_CHI,Tian2023ArXiv,Savelka2023ACM_ITiCSE}, human usability\cite{Zamfirescu2023ACM_CHI,Vaithilingam2022ACM_CHI_EA,Kazemitabaar2023ACM_Koli}, and, yet others have leveraged the LLM architecture for specialized applications \cite{Feng2023arXiv,Kuramitsu2023ACM_SPLASH-E,Pu2023ACM_UIST}.

Many researchers have evaluated the performance of LLMs for specific programming tasks. ChatGPT, for instance, demonstrated proficiency in passing most introductory programming assessments, although struggling with more complex ones \cite{Savelka2023ACM_ITiCSE, Ouh2023ACM_ITiCSE}. Interestingly, when compared to students, LLMs have shown superior performance in these introductory programming assessments \cite{FinnieAinsley2023ACM_ACE}. LLMs have also proven effective in program repair, surpassing the performance of automated program repair techniques \cite{Xia2023ACM_ICSE}. These are just a few examples of the capabilities of LLMs.

In addition to the evaluations done for LLMs against datasets, the usability of LLMs has also gained attention, prompting evaluations involving humans to gauge real-time performance and practical applications \cite{Kim2022ACM_CHI, Weisz2022ACM_IUI, Yen2023arXiv}. While some studies suggest LLMs enhance performance in programming tasks and other contexts, other studies indicated mixed or limited benefits \cite{Madi2023ACM_ASE, Yen2023arXiv}. Several user studies have also delved into the interaction patterns of programmers with LLMs \cite{Barke2023ACM_PL, Vaithilingam2022ACM_CHI_EA, Kazemitabaar2023ACM_Koli}, unveiling diverse patterns of interaction and prompting strategies among programmers.

In this paper, we conducted a literature survey examining user studies assessing the interaction between humans and LLMs, as well as user studies that investigate the human and task performance effects when LLMs are used by humans across various programming tasks. We aimed to analyze LLM performance across diverse tasks and offer practical guidance for non-experts in machine learning (non-AI experts), when utilizing LLMs for programming tasks as well as shed light on the gaps in user studies for researchers to explore. Specifically, our study addressed the following research questions:
\begin{enumerate}
        \item[\textbf{RQ1:}] How are programmers interacting with LLMs (Interaction Observations)?
	\item[\textbf{RQ2:}] Does the use of LLMs lead to enhanced human capabilities (Human Enhancement Evaluation)?
        \item[\textbf{RQ3:}] Does the use of LLMs improve task performance (Task Performance Evaluation)?
        \item[\textbf{RQ4:}] Are there interaction behaviours that lead to enhanced human capabilities or improved task performance (User-LLM Interaction Effects on Human Enhancement and Task Performance)?
\end{enumerate}
By restricting our analysis to observed interactions of the human with the LLM, we gain insight into real-world usage patterns and the effectiveness of LLMs in practical scenarios. This approach provides a more comprehensive understanding of LLMs' capabilities and limitations when utilized by humans.

\section{Research Methodology}
To gather data for our study, we focused on LLMs utilizing the transformers architecture \cite{Vaswani2017NIPS}, widely regarded as the leading framework for LLMs. We conducted a search across various publishers (ACM, IEEE, Scopus and ArXiv), using the following search terms;

\begin{center}
\begin{tabular}{p{1\linewidth}}
\toprule
“ChatGPT” OR “Co-pilot” OR “coPilot” OR “copilot” OR “LLM” OR “LLMs” OR “GPT” OR “large language model” OR “large-language model” OR “large language models” OR “large-language models” \\
\parbox{\linewidth}{\centering \textbf{AND}}\\
“programmers” OR “programmer” \\
\parbox{\linewidth}{\centering \textbf{AND}} \\
“programming” OR “code generation” OR “program synthesis” \\
\bottomrule
\end{tabular}
\end{center}

We intentionally left the search terms to be as broad as possible in order to get a diverse set of papers containing user studies. Despite this, there was still a scarcity of papers. Due to the scarcity of papers, we broadened our scope to encompass experience reports (studies in which the researcher reports on their experience of using an LLM for various tasks), treating them as single-user studies. Our review focused on papers published post-2017, coinciding with the introduction of the transformer architecture. Additionally, we utilized search engine features (where applicable) to refine our results, aiming to restrict them to studies involving human experiments.
 
We examined the title and abstract of each paper to determine if it met all of our inclusion criteria, which are as follows:

\begin{enumerate}
\item The paper pertains to programming with LLMs.
\item The paper includes a user study or is an experience report.
\item The LLM discussed in the paper is based on the transformer architecture.
\end{enumerate}

Additionally, we utilized backward (considering only references published post-2017) and forward snowballing to augment our initial set of papers. 

After delving deeper into the papers, beyond just the abstract, we opted to exclude some papers that were either unrelated to the programming process (i.e. not related to code generation, or code explanations), or did not feature observed interactions of the user with the LLM.
We also had outlined the following exclusion criteria:
\begin{enumerate}
    \item Studies exploring LLM capabilities using datasets
    \item Studies that are off focus from evaluating LLMs or enhancing LLM usage.
\end{enumerate}

\section{General Insights}
After completing our search, we settled on a set of 88 papers. These user studies spanned various fields of computing and encompassed tasks related to both code generation and code analysis. 

We loosely categorized the studies based on objectives and we realized the studies that we found had a diverse set of objectives; some studies were focused on assessing human performance (e.g productivity, pair programming performance, etc), others evaluated the LLM for specific kind of tasks (e.g code quality, data analysis) or domain of tasks (e.g software engineering, data science, etc). We also noticed that our set of papers had applications in computing education and also in computing industry.

\begin{table}[ht]
    \centering
    \begin{tabular}{lcc}
        \toprule
        \textbf{Goal of Study} & \textbf{Number of Papers} \\
        \midrule
        Human Enhancement Evaluation &  8  \\
        Testing or Improving Learning &  17 \\
        Pair Programming &  2 \\
        Creative Coding &  2 \\
        Human-LLM Interaction  &  18 \\
        Evaluating the LLM Response & 10 \\
        Software Engineering Concepts & 16 \\
        Code Security & 6 \\
        Code Translation & 2 \\
        Data Science  & 4 \\
        Embedded Systems & 3 \\
        \bottomrule
    \end{tabular}
    \caption{A very loose categorization of the aim of the user studies evaluated in this survey}
    \vspace{-20pt} % Adjust as needed
    \label{tab:paper_categories}
\end{table}

Among the papers discovered, we found that the human studies conducted could be categorized loosely into two major types of experiments: Controlled/Laboratory experiments and Field Experiments.

Within these types of experimentation, there were 3 data analysis methods that we discovered: human enhancement evaluation, Task performance evaluation and interaction analysis. We illustrate this in Table \ref{tab:paper_summary} to give a summary of the number of papers that focused on each data analysis method. 43 studies used a combination of methods.

\begin{table*}[ht]
    \centering
    \begin{tabular}{p{5cm}p{12cm}}
        \toprule
        \textbf{Analysis Technique} & \textbf{References} \\
        \midrule
        Interaction Analysis (\textbf{RQ1}) 53 papers &  \citet{Vaithilingam2022ACM_CHI_EA}, \citet{Bird2023ACM_Queue}, \citet{Kazemitabaar2023ACM_CHI}, \citet{Qian2024ACM_IUI}, \citet{Tan2024arXiv}, \citet{Sun2024IJETHE}, \citet{Shoufan2023ACM_TCE}, \citet{Liffiton2024ACM_Koli}, \citet{kim2024arXiv}, \citet{Kazemitabaar2024ACM_CHI}, \citet{Sheese2024ACM_ACE}, \citet{Kuramitsu2023ACM_SPLASH-E}, \citet{Jing2024HSSC}, \citet{Yan2024arXiv}, \citet{Vasiliniuc2023IEEE_ICCP}, \citet{Denny2024iTCSE}, \citet{Ouaazki2023TALE}, \citet{Feng2023arXiv}, \citet{Tseng2024arXiv}, \citet{Yin2024arXiv}, \citet{Liu2023ACM_CHI}, \citet{Nam2024ACM_ICSE}, \citet{Rao2024ACM_IUI}, \citet{Chopra2023arXiv1}, \citet{Arawjo2024ACM_CHI}, \citet{Yan2024ACM_CHI}, \citet{Prather2023ACM_TCI}, \citet{Barke2023ACM_PL}, \citet{Kazemitabaar2023ACM_Koli}, \citet{Jiang2022ACM_CHI}, \citet{Mozanner2023rXiv}, \citet{Tang2023CMU}, \citet{Prasad2023ACM_CompEd}, \citet{Arora2024arXiv}, \citet{Nguyen2024ACM_CHI}, \citet{Madi2023ACM_ASE}, \citet{Jayagopal2022ACM_UIST}, \citet{Bernstein2024ITiCSE}, \citet{MacNeil2023ACM_SIGCSE}, \citet{Jury2024ACM_ACE}, \citet{Vaithilingam2024ACM_CHI}, \citet{Ross2023ACM_IUI}, \cite{YM2023arXiv}, \citet{Wang2024arXiv}, \citet{Khojah2024arXiv}, \citet{Liang2024arXiv}, \citet{Patton2024arXiv}, \citet{Perry2023ACM_CCS}, \citet{Oh2023IEEE_SP}, \citet{Feng2024ACM_ICSE}, \citet{Weisz2022ACM_IUI}, \citet{Gu2024ACM_CHI}, \citet{Chopra2023arXiv}, \citet{Wang2024IEEE_TVCG}, \citet{Karli2024ACM_HRI}   \\
        \midrule
        Human Enhancement Evaluation (\textbf{RQ2}) 43 papers &  \citet{Vaithilingam2022ACM_CHI_EA}, \citet{Kazemitabaar2023ACM_CHI}, \citet{Peng2023arXiv}, \citet{Liu2023ArXiv}, \citet{Tan2024arXiv}, \citet{Yilmaz2023CEAI}, \citet{Ma2023arXiv}, \citet{Kosar2024mdpi}, \citet{kim2024arXiv}, \citet{Kuramitsu2023ACM_SPLASH-E}, \citet{Xue2024ACM_ICSE}, \citet{Jing2024HSSC}, \citet{Yan2024arXiv}, \citet{Vasiliniuc2023IEEE_ICCP}, \citet{Denny2024iTCSE}, \citet{Ouaazki2023TALE}, \citet{Imai2022ACM_IEEE}, \citet{Feng2023arXiv}, \citet{Yin2024arXiv}, \citet{Yen2023arXiv}, \citet{Ferdowsi2023arXiv}, \citet{Nam2024ACM_ICSE}, \citet{Rao2024ACM_IUI}, \citet{Chopra2023arXiv1}, \citet{Yan2024ACM_CHI}, \citet{Tang2023CMU}, \citet{Nguyen2024ACM_CHI}, \citet{Fakhoury2024arXiv}, \citet{Vaithilingam2024ACM_CHI}, \citet{Kim2022ACM_CHI}, \citet{Wang2024arXiv}, \citet{Dirin2023rG}, \citet{Chatterjee2024SEC}, \citet{Aillon2023IEEE_C3}, \citet{Sandoval2023USENIX}, \citet{Feng2024ACM_ICSE}, \citet{Su2024ACM_ICSE}, \citet{Weisz2022ACM_IUI}, \citet{Liu2024IEEE_TSE}, \citet{Chopra2023arXiv}, \citet{Johnson2024JRTC}, \citet{Karli2024ACM_HRI}, \\
        \midrule
        Task Performance Evaluation (\textbf{RQ3}) 40 papers & \citet{Kazemitabaar2023ACM_CHI}, \citet{Peng2023arXiv}, \citet{Liu2023ArXiv}, \citet{Qian2024ACM_IUI}, \citet{Tan2024arXiv}, \citet{Sun2024IJETHE}, \citet{Shoufan2023ACM_TCE}, \citet{Xue2024ACM_ICSE}, \citet{Denny2024ACM_SIGCSE}, \citet{Vasiliniuc2023IEEE_ICCP}, \citet{Imai2022ACM_IEEE}, \citet{Ferdowsi2023arXiv}, \citet{Yan2024ACM_CHI}, \citet{Kazemitabaar2023ACM_Koli}, \citet{Jiang2022ACM_CHI}, \citet{Nguyen2024ACM_CHI}, \citet{Fakhoury2024arXiv}, \citet{Madi2023ACM_ASE}, \citet{Nejjar2024ICSSP}, \citet{Wermelinger2023ACM_SIGCSE}, \citet{Cipriano2023ACM_ITiCSE}, \citet{Sanger2024arXiv}, \citet{Erhabor2023arXiv}, \citet{Matthieu2023ACM_SPLC}, \citet{Yiming2023ACM_HOTOS}, \citet{Ross2023ACM_IUI}, \cite{Acher2023SPLC}, \citet{Camara2023SSM}, \citet{Wang2024arXiv}, \citet{Liang2024arXiv}, \citet{Choudhuri2024ACM_ICSE}, \citet{Chatterjee2024SEC}, \citet{Aillon2023IEEE_C3}, \citet{Sandoval2023USENIX}, \citet{Perry2023ACM_CCS}, \citet{Asare2024ACM_ICSE}, \citet{Weisz2022ACM_IUI}, \citet{Zhou2023ACM_CHI_EA}, \citet{Englhardt2023arXiv}, \citet{Johnson2024JRTC},   \\
        \bottomrule
    \end{tabular}
    \caption{Summary of research papers categorized by the analysis methods they employed and the corresponding research questions they addressed.}
    \vspace{-20pt} % Adjust as needed
    \label{tab:paper_summary}
\end{table*}

\subsection{Human Enhancement Evaluation Metrics}
This type of data analysis is an evaluation of the effects that using the LLM has on the user.
We compiled the enhancement metrics from papers assessing user-LLM performance. Despite slight variations in terminology, we grouped metrics with similar names together. There were 2 main human performance metrics found: time productivity and learning check. We discuss these in detail in section \ref{RQ2_Section}. It is important to note that there were more performance metrics assessed within these studies (e.g self-efficacy\cite{Yiming2023ACM_HOTOS, Choudhuri2024ACM_ICSE}, trust, etc). While we recognize that these metrics are also important, we focused here on human performance metrics identified that could be evaluated objectively. 

\subsection{Task Performance Evaluation Metrics}
This type of data analysis is an evaluation of the code produced by the LLM. This includes metrics like code quality, readability, correctness as it relates to the task, etc. We describe these metrics in further detail in section \ref{RQ3_Section}.

\subsection{Human-AI Interaction Data Analysis}
We found that 53 papers evaluated the interactive components of the human with the LLMs. In fact, some studies only focused on this. Within this set of studies, there were 3 major themes encountered when reviewing these papers: 
\begin{enumerate}
    \item The kind of requests made to the LLM 
    \item How the LLM was prompted for a request type
    \item How users interacted with the LLM to accomplish their task
\end{enumerate}
We expand on these themes in section \ref{RQ1_Section}. 

\subsection{Categorization of users}
During our review, we identified two main categorizations: Industry practitioners versus academia users, and novices versus experts. These distinctions provide a reference for assessing the balance of the papers in this survey and the extent to which the user study results can be generalized. 

\subsubsection{Industry Practitioners Vs Academia Users}
This categorization reflects the industry affiliation of participants rather than their expertise in a specific task. Academia users included undergraduates, graduates, and potentially professors (based on experience reports, it’s possible that some papers were authored by professors rather than graduate students). Industry practitioners were those actively working in various fields, not limited to computing. While some academia users may have had industry experience, we primarily relied on the information provided in the studies, avoiding speculation and focusing on participant recruitment sources to form this list. Although the title suggests a comparison, this categorization was not used to differentiate how participants were grouped in the experiments. Instead, it served either as a diversity statement to indicate participant group balance or simply to provide context about participant backgrounds.
We give a detail of the papers associated with each industry in table \ref{tab:user_industry_papers}.

\begin{table*}[ht]
    \centering
    \begin{tabular}{p{5cm}p{12cm}}
    \toprule
    \textbf{User Group} & \textbf{Associated Papers} \\
    \midrule
    Academia Users & \citet{Prather2023ACM_TCI}, \citet{Shoufan2023ACM_TCE}, \citet{Kazemitabaar2023ACM_CHI}, \citet{Tan2024arXiv}, \citet{Sun2024IJETHE}, \citet{Yilmaz2023CEAI}, \citet{Ma2023arXiv}, \citet{Kosar2024mdpi}, \citet{kim2024arXiv}, \citet{Kazemitabaar2024ACM_CHI}, \citet{Liffiton2024ACM_Koli}, \citet{Kuramitsu2023ACM_SPLASH-E}, \citet{Xue2024ACM_ICSE}, \citet{Jing2024HSSC}, \citet{Sheese2024ACM_ACE}, \citet{Denny2024ACM_SIGCSE}, \citet{Yan2024arXiv}, \citet{Denny2024iTCSE}, \citet{Ouaazki2023TALE}, \citet{Imai2022ACM_IEEE}, \citet{Yen2023arXiv}, \citet{Arawjo2024ACM_CHI}, \citet{Yan2024ACM_CHI}, \citet{Kazemitabaar2023ACM_Koli}, \citet{Tang2023CMU}, \citet{Prasad2023ACM_CompEd}, \citet{Arora2024arXiv}, \citet{Nguyen2024ACM_CHI}, \citet{Madi2023ACM_ASE}, \citet{Jayagopal2022ACM_UIST}, \citet{Nejjar2024ICSSP}, \citet{Wermelinger2023ACM_SIGCSE}, \citet{Cipriano2023ACM_ITiCSE}, \citet{Bernstein2024ITiCSE}, \citet{Sanger2024arXiv}, \citet{MacNeil2023ACM_SIGCSE}, \citet{Jury2024ACM_ACE}, \citet{Matthieu2023ACM_SPLC}, \citet{Vaithilingam2024ACM_CHI}, \citet{Acher2023SPLC}, \citet{YM2023arXiv}, \citet{Camara2023SSM}, \citet{Wang2024arXiv}, \citet{Liang2024arXiv}, \citet{Choudhuri2024ACM_ICSE}, \citet{Dirin2023rG}, \citet{Patton2024arXiv}, \citet{Aillon2023IEEE_C3}, \citet{Sandoval2023USENIX}, \citet{Feng2024ACM_ICSE}, \citet{Zhou2023ACM_CHI_EA}, \citet{Gu2024ACM_CHI}, \citet{Johnson2024JRTC}, \citet{Karli2024ACM_HRI},  \\
    \midrule
     Industry Practitioners & \citet{Vaithilingam2022ACM_CHI_EA}(1user), \citet{Peng2023arXiv}, \citet{Qian2024ACM_IUI}, \citet{Vasiliniuc2023IEEE_ICCP}, \citet{Tseng2024arXiv}, \citet{Yin2024arXiv}, \citet{Liu2023ACM_CHI}, \citet{Rao2024ACM_IUI}, \citet{Chopra2023arXiv1}, \citet{Jiang2022ACM_CHI}, \citet{Mozanner2023rXiv}, \citet{Yiming2023ACM_HOTOS}, \citet{Ross2023ACM_IUI}, \citet{Khojah2024arXiv}, \citet{Chatterjee2024SEC}, \citet{Weisz2022ACM_IUI}, \citet{Liu2024IEEE_TSE}, \citet{Chopra2023arXiv}, \citet{Wang2024IEEE_TVCG} \\
     \midrule
    Mixed Industries & \citet{Feng2023arXiv}, \citet{Ferdowsi2023arXiv}, \citet{Nam2024ACM_ICSE}, \citet{Barke2023ACM_PL}, \citet{Fakhoury2024arXiv}, \citet{Erhabor2023arXiv}, \citet{Perry2023ACM_CCS}, \citet{Oh2023IEEE_SP}, \citet{Asare2024ACM_ICSE}, \citet{Su2024ACM_ICSE}, \citet{Englhardt2023arXiv},  \\
    \bottomrule
    \end{tabular}
    \caption{Categorization of user groups based on their industry affiliations.}
    \label{tab:user_industry_papers}
    \vspace{-10pt} % Adjust as needed
\end{table*}

\subsubsection{Novices Vs Experts}
This categorization is based on the users' familiarity with the task. In the evaluated user studies, participants were classified as novices or experts according to their expertise in the task domain. Each study had its own criteria for defining expertise, which could be based on programming language knowledge, task-specific experience, or a combination of both. We adhered to the expertise assessments specified in each paper, focusing on how they related to the task being evaluated.

Notably, we observed that even among groups categorized as "experts" or "novices," there existed variations in expertise levels. To account for this and to provide clearer terminology regarding expertise levels, we adopted the Dreyfus Model of Skill Acquisition \cite{Dreyfus1980}, although with slight modifications \cite{Seth2010}. We categorized "novice" programmers as "novices/beginners" and "expert" programmers as "Competent/Proficient/Experts," aiming to account for skill level variations within the same group. As an abbreviation, we still stick to the terms "novices" and "experts" when referencing the participant pools in this paper.

\begin{table}[ht]
    \centering
    \begin{tabular}{ll}
    \toprule
    Participant Pool Groups & Dreyfus Categories \\
    \midrule
    "Novice" & Novice/Beginner\\
    "Expert" & Competent/Proficient/Expert \\
    "Mixed Group" & All Groups stated above \\
    \bottomrule
    \end{tabular}
    \caption{Participant Groups in Studies and Dreyfus Model Categories}
    \label{tab:Exp_level}
    \vspace{-10pt} % Adjust as needed
\end{table}

For single-user studies (experience reports), we grouped participants as experts based on the assumption that some level of expertise is required to write a research paper. For users studies using students taking a course, those users were categorized as novices (unless stated otherwise). We believe this is a reasonable assumption as students taking an undergraduate course are doing it because of a lack of expertise in the subject of study. 
27 studies also decided to used mixed groups, without assessing how expertise affected performance.

We give a list of the papers categorized by expertise level in Table  \ref{tab:user_exp_papers}.

\begin{table*}[ht]
    \centering
    \begin{tabular}{p{5cm}p{12cm}}
    \toprule
    \textbf{User Group} & \textbf{Associated Papers} \\
    \midrule
    Novices & \citet{Prather2023ACM_TCI}, \citet{Shoufan2023ACM_TCE}, \citet{Kazemitabaar2023ACM_CHI}, \citet{Sun2024IJETHE}, \citet{Yilmaz2023CEAI}, \citet{Ma2023arXiv}, \citet{Kosar2024mdpi}, \citet{Kazemitabaar2024ACM_CHI}, \citet{Liffiton2024ACM_Koli}, \citet{Kuramitsu2023ACM_SPLASH-E}, \citet{Kuramitsu2023ACM_SPLASH-E}, \citet{Denny2024ACM_SIGCSE}, \citet{Yan2024arXiv}, \citet{Vasiliniuc2023IEEE_ICCP}, \citet{Denny2024iTCSE}, \citet{Ouaazki2023TALE}, \citet{Imai2022ACM_IEEE}, \citet{Kazemitabaar2023ACM_Koli}, \citet{Prasad2023ACM_CompEd}, \citet{Nguyen2024ACM_CHI}, \citet{Jayagopal2022ACM_UIST}, \citet{Bernstein2024ITiCSE}, \citet{MacNeil2023ACM_SIGCSE}, \citet{Jury2024ACM_ACE}, \citet{Kim2022ACM_CHI}, \citet{Choudhuri2024ACM_ICSE}, \citet{Dirin2023rG}, \citet{Johnson2024JRTC},  \\
    \midrule
    Experts & \citet{Liu2023ArXiv}, \citet{Qian2024ACM_IUI}, \citet{Tan2024arXiv}, \citet{Xiao2024ACM_CHI}, \citet{Feng2023arXiv}, \citet{Yen2023arXiv}, \citet{Barke2023ACM_PL}, \citet{Mozanner2023rXiv}, \citet{Fakhoury2024arXiv}, \citet{Nejjar2024ICSSP}, \citet{Wermelinger2023ACM_SIGCSE}, \citet{Cipriano2023ACM_ITiCSE}, \citet{Sanger2024arXiv}, \citet{Erhabor2023arXiv}, \citet{Matthieu2023ACM_SPLC}, \citet{Vaithilingam2024ACM_CHI}, \citet{Yiming2023ACM_HOTOS}, \citet{Acher2023SPLC}, \citet{Camara2023SSM}, \citet{Wang2024arXiv}, \citet{Khojah2024arXiv}, \citet{Liang2024arXiv}, \citet{Chatterjee2024SEC}, \citet{Aillon2023IEEE_C3}, \citet{Asare2024ACM_ICSE}, \citet{Feng2024ACM_ICSE}, \citet{Liu2024IEEE_TSE}, \citet{Zhou2023ACM_CHI_EA}, \citet{Gu2024ACM_CHI}, \citet{Chopra2023arXiv},   \\
    \midrule
    Mixed Group & \citet{Vaithilingam2022ACM_CHI_EA}, \citet{kim2024arXiv}, \citet{Jing2024HSSC}, \citet{Tseng2024arXiv}, \citet{Yin2024arXiv}, \citet{Liu2023ACM_CHI}, \citet{Ferdowsi2023arXiv}, \citet{Nam2024ACM_ICSE}, \citet{Rao2024ACM_IUI}, \citet{Chopra2023arXiv1}, \citet{Arawjo2024ACM_CHI}, \citet{Yan2024ACM_CHI}, \citet{Jiang2022ACM_CHI}, \citet{Tang2023CMU}, \citet{Arora2024arXiv}, \citet{Madi2023ACM_ASE}, \citet{Ross2023ACM_IUI}, \citet{YM2023arXiv}, \citet{Patton2024arXiv}, \citet{Sandoval2023USENIX}, \citet{Perry2023ACM_CCS}, \citet{Oh2023IEEE_SP}, \citet{Su2024ACM_ICSE}, \citet{Weisz2022ACM_IUI}, \citet{Wang2024IEEE_TVCG}, \citet{Englhardt2023arXiv}, \citet{Karli2024ACM_HRI},  \\
    \bottomrule
    \end{tabular}
    \caption{User Groups by familiarity with tasks along with the papers that report on this information on participants}
    \label{tab:user_exp_papers}
    \vspace{-10pt} % Adjust as needed
\end{table*}

\section{RQ1: Interaction Observations}
\label{RQ1_Section}
\subsubsection{How we addressed it}
We qualitatively answer this research question by examining studies that investigated the interactive component of users and LLMs and collated their findings.

\subsubsection{The Results}
As mentioned in the previous section; there were 3 main themes identified when reviewing the user-LLM interaction data reported in the studies. In this section, we explain them in further detail.

\subsection{The kind of requests made to the LLM} 
\label{request_types}
This category has to do with the type of requests users made of the LLM. 
% This can also be described as phases of conversations with the LLM \cite{Barke2023ACM_PL, Arawjo2024ACM_CHI} which were outlined by Mozanner et al. \cite{Mozanner2023rXiv} in a lot more detail. 
A number of studies reported upon this and upon synthesizing all of their reports, we discovered that users made 4 types of inquiries when requesting information from the LLM. In the following explanations, there may be some confusion regarding whether this categorization pertains to individual prompts or different prompting phases. We observed that studies focusing on interaction analysis (beyond just examining the prompts) describe these categories as phases of interaction, while those concentrating on prompt analysis refer to them as types of requests. Through our analysis and synthesis, we found that these concepts are closely related, so we have combined them in our discussion below.

\subsubsection{Learning/Exploration Requests:} This category/phase includes requests where the user requests information from the LLM in order to gain an understanding of the task/topic/code \cite{Sheese2024ACM_ACE, Ross2023ACM_IUI, Kuramitsu2023ACM_SPLASH-E, Perry2023ACM_CCS, Arora2024arXiv, Ouaazki2023TALE, Xue2024ACM_ICSE, Kazemitabaar2024ACM_CHI, Khojah2024arXiv, Yin2024arXiv, Barke2023ACM_PL, Vasiliniuc2023IEEE_ICCP, Prasad2023ACM_CompEd} or exploring ways to implement the task \cite{Rao2024ACM_IUI, Arora2024arXiv, Khojah2024arXiv}. With these kind of prompts, users were using the LLM as an expert consultant \cite{Khojah2024arXiv} as well as for personalized learning. This form of interaction was shown to involve many steps \cite{Chopra2023arXiv}. In two cases, a user interface (UI) that involved interacting with the LLM beyond the typical chat interface, this type of request was still observed \cite{Rao2024ACM_IUI, Khojah2024arXiv}. 
While not a large percentage of prompts belong to this category \cite{Prasad2023ACM_CompEd, Arora2024arXiv} (out of the total number of prompts), there were notable cases where users continued prompting the LLM with exploration prompts \cite{Yin2024arXiv, Arawjo2024ACM_CHI} for the entirety of the task time (for user studies where a specific task was given). This was noted to occur among novice users. It is unclear if the exploration prompting led participants to not complete the task or if they just inquired from the LLM and proceeded to solve the task themselves.
One study made note of the fact that the purpose for which users used this sort of request was achieved as they were able to learn something \cite{Xue2024ACM_ICSE}. We can find more insight on whether users are learning with the use of LLMs in our evaluation of human enhancement metrics in  section \ref{RQ2_Section}.

\subsubsection{Solution-oriented Prompts:} This category/phase includes requests that have to do with generating information specific to the task. A large percentage of prompts were found to belong to this category \cite{Bernstein2024ITiCSE, Khojah2024arXiv}. Users requests in this case include natural language code implementation details \cite{Sheese2024ACM_ACE, Bernstein2024ITiCSE, Perry2023ACM_CCS, Xue2024ACM_ICSE, Kuramitsu2023ACM_SPLASH-E, Prasad2023ACM_CompEd, Tseng2024arXiv}, test case generation \cite{Arora2024arXiv}, requests to improve on current code \cite{Khojah2024arXiv, Perry2023ACM_CCS, Ouaazki2023TALE}, code documentation \cite{Arora2024arXiv}, and, pseudocode prompts \cite{Kazemitabaar2023ACM_Koli, Perry2023ACM_CCS}. These implementation queries varied in terms of details with some being quite broad and others being very specific\cite{Tseng2024arXiv}.
With these type of requests, there was some indication of a copying and pasting behaviour where the similarity to the task description was fairly high \cite{Perry2023ACM_CCS, Arora2024arXiv, Kazemitabaar2023ACM_Koli, Denny2024iTCSE}. This behaviour was observed among studies that made use of students (novices) as participants.

\subsubsection{Error-Correcting Prompts:} This category includes requests where the user requested that the LLM debug their code \cite{Sheese2024ACM_ACE, Arora2024arXiv, Kazemitabaar2024ACM_CHI, Wang2024arXiv} or they reported an error in the response. In the case of reporting an incorrect response, there were situations where users did not give details about how the LLM should fix the response \cite{Sheese2024ACM_ACE} and there were instances where users would tell the LLM clearly what the issue is and how to correct the error by refining their prompts \cite{Ouaazki2023TALE, Kuramitsu2023ACM_SPLASH-E, Yan2024ACM_CHI}.

\subsubsection{Unrelated to solution Prompts:} There were requests where users created prompts unrelated to the task. These were relational prompts \cite{Kuramitsu2023ACM_SPLASH-E, kim2024arXiv} such as thanking the LLM as well as cases where the context did not align with task at hand and thus could not be categorized properly by the researchers \cite{Arora2024arXiv, Sheese2024ACM_ACE, Prasad2023ACM_CompEd}.

\subsection{How the LLM was prompted for a request type}
\label{prompting_patterns}
This category examines the prompting strategies used by users to elicit desired responses from the LLM. We excluded any prompting patterns identified in experience reports, focusing instead on how programmers without much experience in prompting patterns approach the task of prompting.

With regards to how participants started their prompting interaction with the LLM, there were two main prompting styles.
\begin{itemize}
    \item \textbf{Single Prompt Method:} This can also be referred to as the the zero-shot method (to borrow from machine learning terminology). This was a prompting method where users requested a solution to their issue from the LLM all in one prompt \cite{Ross2023ACM_IUI, Kazemitabaar2023ACM_Koli, Tseng2024arXiv, Jiang2022ACM_CHI}. 
    \item \textbf{Multi-Prompt Method:} This was a method where users would initially break down tasks (intentionally or unintentionally) and request for a solution to each sub-task in a separate prompt \cite{Ross2023ACM_IUI, Kazemitabaar2023ACM_CHI, Kazemitabaar2023ACM_Koli}.
\end{itemize}

Some users initially employed the single prompt method \cite{Kazemitabaar2023ACM_CHI, Zhou2023ACM_CHI_EA, Ross2023ACM_IUI}, while others used the multi-prompt method \cite{Zhou2023ACM_CHI_EA, Ross2023ACM_IUI, Kazemitabaar2023ACM_Koli, Englhardt2023arXiv}.

This initial categorization of LLM prompting helps us understand how users start out prompting the LLM. In the course of the interaction, users incorporated re-prompting strategies in order to get back on track when the LLM failed to produce the correct response. Note that these prompting patterns we identified from synthesizing the studies are not context-specific and they also don't encompass the entire user task. They generally address the methods that users used to get at a particular solution needed from the LLM, whether it be an implementation solution, debugging issue or learning prompt. These patterns we outline below:

\subsubsection{Re-asking the Question:} This was a prompting pattern in which users tried to get the LLM to regenerate the response by just asking the exact question again\cite{Shoufan2023ACM_TCE}. It was identified among novices. Due to the non-deterministic nature of LLMs, this seems to be a valid approach.

\subsubsection{Asking the LLM for verification:} This pattern was also identified among novices \cite{Shoufan2023ACM_TCE}. Users, in an attempt to verify the LLM response would ask the LLM to double-check it's own response. According to the study done by \citet{Shoufan2023ACM_TCE}, it seemed to be used to getting reassurance from the LLM as well as getting the LLM to double-check its response.

\subsubsection{Task Transformation:} This was a case where participants out of frustration or some other reason, decide to completely change the nature of the task that they wanted the LLM to solve\cite{Liu2023ACM_CHI, Yin2024arXiv}. This method is usually used after several attempts at trying to get the LLM to do something and it failing repetitively. 

\subsubsection{Word Restructuring:} With this method, users resorted to rephrasing their prompts in order to get the LLM to elicit the right response \cite{Liu2023ACM_CHI, Yan2024arXiv}.

\subsubsection{Elaborating on a task:} This was a prompting pattern where users expanded on a prompt by providing more context or specific details regarding the task. This was the most commonly identified prompting pattern among the studies evaluated in this paper (most likely because of its intuitive nature). It was a prompting pattern identified among novices and experts \cite{Kazemitabaar2023ACM_Koli, Englhardt2023arXiv}.
Users used this strategy by either breaking down the task to be performed into steps of what to do (a form of multi-prompting) \cite{Nam2024ACM_ICSE, Yin2024arXiv, Tseng2024arXiv, Liu2023ACM_CHI, Denny2024iTCSE, Kazemitabaar2023ACM_Koli, Barke2023ACM_PL, Arawjo2024ACM_CHI, Vasiliniuc2023IEEE_ICCP} or by providing general high-level context regarding the kind of output the user wanted to see (a form of single prompting) \cite{Yin2024arXiv, Ouaazki2023TALE, Nam2024ACM_ICSE, Denny2024iTCSE, Liu2023ACM_CHI, Yan2024arXiv}. 

As expected, when users elaborated on the task, their prompts were much longer \cite{Bernstein2024ITiCSE} as expected. As noted in a few studies, many users struggled with providing the LLM with the context it needed and this was identified among novices\cite{Denny2024iTCSE, Sheese2024ACM_ACE, Kazemitabaar2023ACM_Koli} and experts \cite{Khojah2024arXiv, Chopra2023arXiv} and in studies with mixed participants \cite{Tseng2024arXiv}. Since the struggle for providing context for the LLM seems to be universal, there is a need for tools that guide users towards providing more context.

\subsubsection{Reducing Scope:} In this case, users re-prompted the LLM by reducing the scope of the task to be performed\cite{Liu2023ACM_CHI}. This could imply that they resorted to executing certain parts of the task themselves.

\subsubsection{Combining Prompting patterns}: Another thing that we identified from our analysis is that users used a combination of techniques to arrive at their solution \cite{Yin2024arXiv}. This makes sense and is also inline with the non-deterministic nature of LLMs. Users will use whatever methods they can to arrive at their solution. There was one instance \cite{Yin2024arXiv}, where users went through different prompting patterns very quickly. This could be an indication of a poor understanding of the task at hand.

\subsection{How users interacted with the LLM to accomplish their task}
This section deals with observations of users as they interact with the LLM. This covers other aspects of the interaction beyond prompting the LLM and investigates how users accomplish the given task when given the choice of using an LLM.

\subsubsection{Nature of LLM Interaction Observations:} This category has to do with whether users incorporated other tools into their use of LLMs and how they did it.
The nature of the LLM interaction fell into 3 categories; Only LLM interaction, Hybrid LLM interaction and no LLM interaction.

\textbf{Only LLM Interaction:} This was a situation where users prompted the LLM for the entire solution\cite{Jiang2022ACM_CHI, Karli2024ACM_HRI} and refrained from doing any part of the task manually. Users seem to use this strategy if the LLM response was sufficient \cite{Jury2024ACM_ACE, Jiang2022ACM_CHI} or in cases where they do not know how to do the task on their own. This type of interaction can indicate a dependence on the LLM \cite{Kazemitabaar2023ACM_Koli, Qian2024ACM_IUI}. This dependence on the LLM was discovered in primarily two ways:
\begin{enumerate}
    \item Similarity between final LLM output and what users submitted as their response to the task \cite{Xue2024ACM_ICSE, Kazemitabaar2023ACM_Koli}.
    \item The time users spent reviewing the LLM response \cite{Madi2023ACM_ASE, Yan2024ACM_CHI}.
\end{enumerate}

\textbf{No LLM interaction:} With this type of interaction, users would solve the task completely on their own and would not opt for other resources \cite{Kazemitabaar2023ACM_Koli}. There were instances where participants opted for regular programming \cite{Kazemitabaar2023ACM_Koli, Jiang2022ACM_CHI}, potentially as a strategy to alleviate the cognitive burden of understanding LLM-generated code. 
This type of interaction was rarely found among studies probably because of the inherent nature of the experiments \cite{Liffiton2024ACM_Koli}; if you allow someone an opportunity to make their life easier, at the very least, they would try to use it. Hence, why this interaction method is rare.

\textbf{Hybrid LLM Interaction}: This was where programmers opted for using other resources alongside the LLM \cite{Liang2024arXiv, Rao2024ACM_IUI, Kazemitabaar2023ACM_Koli, Tseng2024arXiv, Mozanner2023rXiv, Yen2023arXiv}. 
There were also hybrid interaction cases where the interface by which users were interacting with the LLM was hybrid giving users the freedom of doing the task manually or using LLM prompting. Some users used more Natural Language (NL) programming while in other cases, more of regular programming was done \cite{Rao2024ACM_IUI, Nam2024ACM_ICSE}. There was also some indication that experts tend to do more regular programming than novices \cite{Qian2024ACM_IUI, Jiang2022ACM_CHI}. 
There were 3 reasons identified as to why users may opt for this type of interaction:
\begin{enumerate}
    \item To alleviate some of the cognitive burden of reviewing LLM response (most common for code generation cases) \cite{Kazemitabaar2023ACM_Koli, Jiang2022ACM_CHI, Patton2024arXiv, Vaithilingam2022ACM_CHI_EA, Yiming2023ACM_HOTOS}.
    \item User already knows what they want to do and thus only needs LLM to generate response for certain components of the task which led to time efficiency \cite{Tseng2024arXiv, Feng2023arXiv}.
    \item It could just be a preference to want to do some LLM interaction as well as manual work \cite{Nam2024ACM_ICSE,Jayagopal2022ACM_UIST}. For example, in cases where the user desires to learn rather than just getting a response.
\end{enumerate}

\subsubsection{Task Time Allocation Observations:}
This pertains to how users manage their time across different stages of interaction with the LLM. Users engage in four main steps when interacting with LLMs: planning and prompt crafting, understanding LLM responses, making changes to LLM code, and accepting or rejecting the LLM response. \citet{Mozanner2023rXiv} provided a comprehensive taxonomy of the programmer's activities during LLM interaction, from which we derived these steps for brevity. The phases of interaction identified by \citet{Mozanner2023rXiv} does have some overlap with the types of requests made to the LLM.

Across all studies, more time was spent in the planning and prompt crafting phases and in the understanding LLM responses phases, with most studies indicating that users were spending more than 50\% of their time in the prompt crafting and understanding phases\cite{Tang2023CMU, Mozanner2023rXiv, Chopra2023arXiv}. Sometimes, this was based on the length of the LLM response \cite{MacNeil2023ACM_SIGCSE}. Out of these two phases of interaction, it seems more time was spent in understanding LLM responses rather than in the prompt crafting phase \cite{Barke2023ACM_PL, Bird2023ACM_Queue, Weisz2022ACM_IUI, Mozanner2023rXiv, Tang2023CMU}. This is inline with the understanding that the introduction of LLMs has caused users to shift from code writing to more code reviewing. Additionally, there were instances where participants either skipped understanding the LLM response entirely or deferred it to a later stage \cite{Vaithilingam2022ACM_CHI_EA, Mozanner2023rXiv}.

\subsubsection{LLM response Review behaviours:}
This theme has to do with the way users reviewed the LLM response. With regular programming, we spend time reading and writing but in the era of LLMs, our time is spent more on reading/reviewing the LLM's response \cite{Sun2024IJETHE}. From the previous sections (\ref{prompting_patterns}, \ref{request_types}), we could infer that reviewing of LLM response is an important process in human-LLM interaction, hence why we have this section. 
Through our synthesis from the papers, we observed 4 LLM response review behaviour types:
\begin{enumerate}
    \item Doing it manually: With this method, users opted for manually making edits to the LLM response in order to test it. Some of these behaviours include; playing around with the variables, temporarily removing the code to test it's effectiveness, editing syntax, etc \cite{Kazemitabaar2023ACM_Koli}.
    \item Doing it via further prompting: Users would work with the LLM to figure out what was wrong \cite{Tang2023CMU, Karli2024ACM_HRI}. In response to this kind of method, a tool called Robin was introduced to help with this \cite{Chopra2023arXiv1}.
    \item Deferring it for later: In this behaviour pattern, users opted to do the verification process later \cite{Vaithilingam2022ACM_CHI_EA, Prather2023ACM_TCI, Mozanner2023rXiv, Xue2024ACM_ICSE}. In these cases, they would just accept the code initially and may later on reject it if discovered to be incorrect \cite{Prather2023ACM_TCI}.
    \item No Review at all: In this case, users skipped review of the LLM response all together \cite{Kazemitabaar2023ACM_Koli}. It was observed in a study with experts that this rarely happens \cite{Jiang2022ACM_CHI}.
\end{enumerate}

There are 3 factors that we observed through our synthesis that informed what LLM response review behaviour that users chose to use.
\begin{enumerate}
    \item Proficiency in task: Users who had more experience with the task at hand were less likely to accept LLM code \cite{Tan2024arXiv, Mozanner2023rXiv, Jing2024HSSC}. There are 2 reasons why this could be happening; the LLM's response is not up to par (maybe due to insufficient information in prompt) or users with more expertise just have a set plan on what they wanted to do and if the LLM defers, they would just reject the LLM response. It is also important to note that there was an instance where proficiency in task did not yield positive gains for the task \cite{Oh2023IEEE_SP}.
    \item Length of LLM response: The length of the LLM response does play a strong role in whether users accepted it upfront \cite{Prather2023ACM_TCI, Mozanner2023rXiv, Feng2024ACM_ICSE, MacNeil2023ACM_SIGCSE, Tan2024arXiv}. A larger LLM response takes more time to review and it makes sense that users are less likely to want to review the LLM's response thoroughly.
    \item Usefulness of Suggestion: This has to do with if the LLM response is in line with the users needs. For example, in one study; code completion tools where strings and comments were suggested, were accepted less by users \cite{Tan2024arXiv}.
\end{enumerate}

With these factors, we identified 3 high level suggestions for reducing the cognitive burden of LLM response review;
\begin{enumerate}
    \item Well-timed and relevant suggestions: LLM creators (as well as interface designers) should make sure that the LLM response is shown at the right time and is relevant to the task at hand for the user \cite{Gu2024ACM_CHI}.
    \item Segmenting the LLM response: By doing this, users are able to take in as much as they can per time and are not overwhelmed \cite{Tan2024arXiv}. In one study by MacNeil et al.\cite{MacNeil2023ACM_SIGCSE}, they found that code explanations that were separated line by line were viewed more by novice users.
\end{enumerate}

\begin{tcolorbox}[colback=gray!10]
\textbf{Summary of RQ1:}\\
We observed several key aspects of users' interactions with LLMs. First, users have specific types of requests and employ diverse prompting strategies to address them. Depending on the perceived usefulness of the LLM responses, users might choose to utilize additional resources or abandon LLM prompting altogether if it proves unproductive. During their interactions, a significant amount of time is spent reviewing responses, and users have developed nuanced behaviors to manage this meticulous task of reviewing LLM responses.
\end{tcolorbox}

\section{RQ2: Human Enhancement Evaluation}
\label{RQ2_Section}
\subsubsection{How we addressed it}
To address this research question, our focus was on discovering and synthesizing the performance metrics associated with the human. We grouped similar metrics together and talk about them under one theme (See Table \ref{tab:human_performance}).

\begin{table*}[ht]
    \centering
    \begin{tabular}{llc}
        \toprule
        \textbf{Metric Theme} & \textbf{Method of evaluation used in Paper} & \textbf{Number of Papers}\\
        \midrule
        \multirow{5}{*}{Time Productivity} & Number of Tasks Completed & 9 \\
                                     & Task completion Time & 18\\
                                     & Lines of Code Written & 1\\
                                     & Number of Errors Encountered & 4\\
                                     & Productivity & 1\\
        \midrule
        \multirow{4}{*}{Learning Check} & Pre-test and Post-test Scores & 7 \\
                                     & Pre-test and Post-test Completion Time & 1 \\
                                     & Course Performance between groups (as Post-test) & 2 \\
                                     & Computational thinking measure (Pre and Post tests) & 2 \\
                                     % & Learning by doing & 2 \\
        \bottomrule
    \end{tabular}
    \caption{The human enhancement themes discovered from the user studies along with the various methods of evaluation used within the papers and the number of papers that used a specific metric.}
    \label{tab:human_performance}
\end{table*}

\subsubsection{The Results}
Overall, significant improvements were observed across various evaluation metrics, with few instances of notable deterioration. 

\textbf{Time Productivity:} This metric is an assessment on how much users were able to accomplish in a given time period. We discovered that this was the most studied human performance metric. Across all studies, significant gains were discovered in the amount of tasks users were able to complete \cite{Kazemitabaar2023ACM_CHI, Tan2024arXiv, Su2024ACM_ICSE, Denny2024iTCSE, Yan2024ACM_CHI, Vasiliniuc2023IEEE_ICCP, Dirin2023rG}, the amount of time it took them to complete each task \cite{Kazemitabaar2023ACM_CHI, Peng2023arXiv, Liu2023ArXiv, Chatterjee2024SEC, Wang2024arXiv, Kim2022ACM_CHI, Weisz2022ACM_IUI}, the amount of code they wrote \cite{Sandoval2023USENIX}, pair programming task time \cite{Imai2022ACM_IEEE, Feng2023arXiv}. The number of errors encountered per task were also much less \cite{Chatterjee2024SEC, Kuramitsu2023ACM_SPLASH-E, Chopra2023arXiv1, Kazemitabaar2023ACM_CHI}. This is an indication that LLMs are good at code syntax.

There were instances where the results of the time evaluation was not significant \cite{Vaithilingam2022ACM_CHI_EA, Liu2023ArXiv, Karli2024ACM_HRI, Xue2024ACM_ICSE, Aillon2023IEEE_C3, Vaithilingam2024ACM_CHI, Weisz2022ACM_IUI, Ferdowsi2023arXiv, Fakhoury2024arXiv, Yan2024arXiv}, as well as instances where the use of LLMs had an opposite effect on time productivity. We discovered that this opposite effect was due to the complexity of the task \cite{Chopra2023arXiv, Nguyen2024ACM_CHI, Yin2024arXiv, Tang2023CMU, Liu2024IEEE_TSE, Yen2023arXiv} or the nature of participants expertise on the task \cite{Rao2024ACM_IUI, Nam2024ACM_ICSE, Kim2022ACM_CHI, Tan2024arXiv} which resulted in users spending more time trying to understand, comprehend and review the LLM's response. This negative effect on time was also seen in our synthesis of the interaction data (See Section \ref{RQ1_Section}).
It is important to note as well that the results of each study when assessing time productivity was not all on one side or the other, a few of the studies had mixed results \cite{Liu2023ArXiv, Fakhoury2024arXiv, Yan2024arXiv, Weisz2022ACM_IUI, Dirin2023rG, Tan2024arXiv, Chopra2023arXiv}.

\textbf{Learning Check}: The metric is an assessment of how much and what students had learnt during their LLM interaction. There seems to be a difference in results based on the test design. When tested using a pre-test and post-test scores design (i.e. users were tested before use of LLM and then tested after use of LLM and score difference was calculated), significant gains were observed between LLM use group and no-LLM groups\cite{Yilmaz2023CEAI, Ma2023arXiv, Ouaazki2023TALE}. It is worth noting that the positive learning outcomes in these studies tended to come from structured or guided LLM implementations - for instance, \citet{Ma2023arXiv} used a purpose-built LLM-based tutoring system - rather than unstructured access to a general-purpose LLM. This suggests that implementation structure may be an important moderating variable for learning outcomes, a distinction we elaborate on further in Section~\ref{sec:llms-for-learning}. Additionally, \citet{Kazemitabaar2023ACM_CHI} found significant gains during the training phase on task performance, but the post-test learning retention result one week later did not reach statistical significance. In cases where students were given access to the LLM during the course/task, no difference in final grades/learning outcomes was observed \cite{kim2024arXiv, Kosar2024mdpi, Xue2024ACM_ICSE}. In yet another study, where the pre-test and post-test design was used but the post-test accessed ability to write the code manually, the group without LLM use performed better \cite{Johnson2024JRTC}. All of the other studies tested on a variety of things: code comprehension and maybe some code writing but the results of this study shows the negative effects of using LLMs on learning proper programming syntax. Some might argue that the learning of programming syntax is becoming obsolete.

\begin{figure*}[h]
  \centering
  \includegraphics[width=0.8\textwidth]{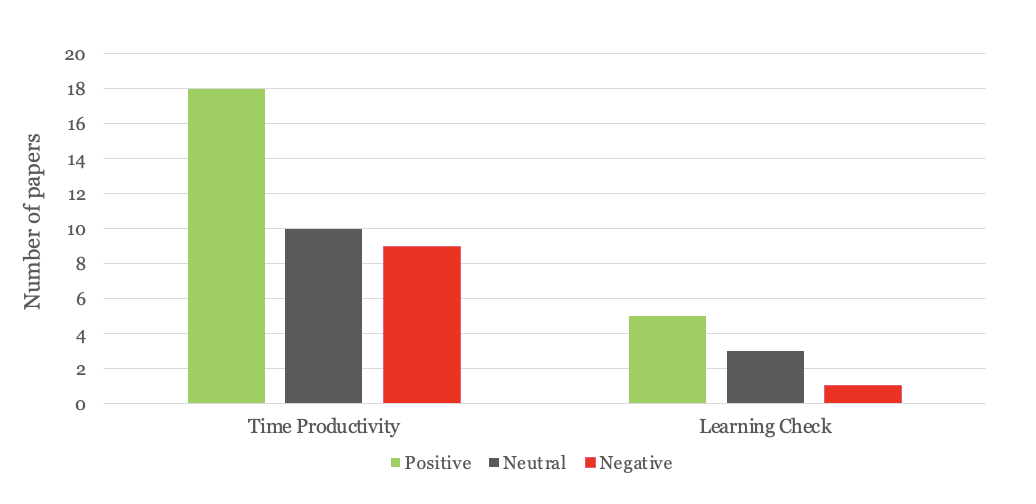}
  \caption{Human enhancement themes categorized by the number of papers reporting positive, neutral, and negative effects.}
  \label{fig:human_enhancement_results}
\end{figure*}

\begin{tcolorbox}[colback=gray!10]
\textbf{Summary of RQ2}\\
Our synthesis reveals mixed results — LLMs can significantly improve time productivity and learning in certain contexts, but a substantial portion of studies found no significant improvement or negative effects. Results vary considerably depending on task type, implementation approach, and user expertise, suggesting these are important moderating factors. In particular, positive learning outcomes appear to be more consistent in structured or guided LLM implementations than in unstructured access scenarios.
\end{tcolorbox}

\section{RQ3: Task Performance Evaluation}
\label{RQ3_Section}
\subsubsection{How we addressed it}
To address this research question, our focus was on evaluating the final submitted human-LLM response at the end of a user session. Specifically, we focused on retrieving the LLM evaluation metrics used in the studies which include metrics like code quality, readability, program speed, etc. We also grouped these metrics into major themes for the sake of brevity (See \ref{tab:code_evaluation}.

\begin{table*}[ht]
    \centering
    \begin{tabular}{llc}
        \toprule
        \textbf{Metric Theme} & \textbf{Method of Evaluation used in Paper} & \textbf{Number of Papers}\\
        \midrule
        \multirow{7}{*}{Collaborative Correctness} & Task-Specific Accuracy Score & 24 \\
                                     & Success Rate  & 4\\
                                     & Unit Tests & 2\\
                                     & Rubric scores & 1\\
                                     & Quality of Solution & 3\\
                                     & Accuracy of Syntax & 2 \\
                                     & Deleted Line counts & 1 \\
        \midrule
        
        \multirow{1}{*}{Code Security} & Code Smells/Bugs & 6 \\        
        \midrule
        
        \multirow{3}{*}{Readability} & Halstead measures of Readability & 1 \\
                                     & Comprehensibility of Code & 1 \\
                                     & Complexity of Code & 1 \\
        \midrule

        Program Speed & Program Speed & 1\\
        \bottomrule
    \end{tabular}
    \caption{The task performance evaluation themes discovered from the studies along with the various methods of evaluation used within papers and the number of papers that used a specific metric}
    \label{tab:code_evaluation}
\end{table*}

\subsubsection{The Results}
We found a number of metrics used to evaluate the human-LLM response. We discuss them below: 

\begin{figure*}[h]
  \centering
  \includegraphics[width=0.8\textwidth]{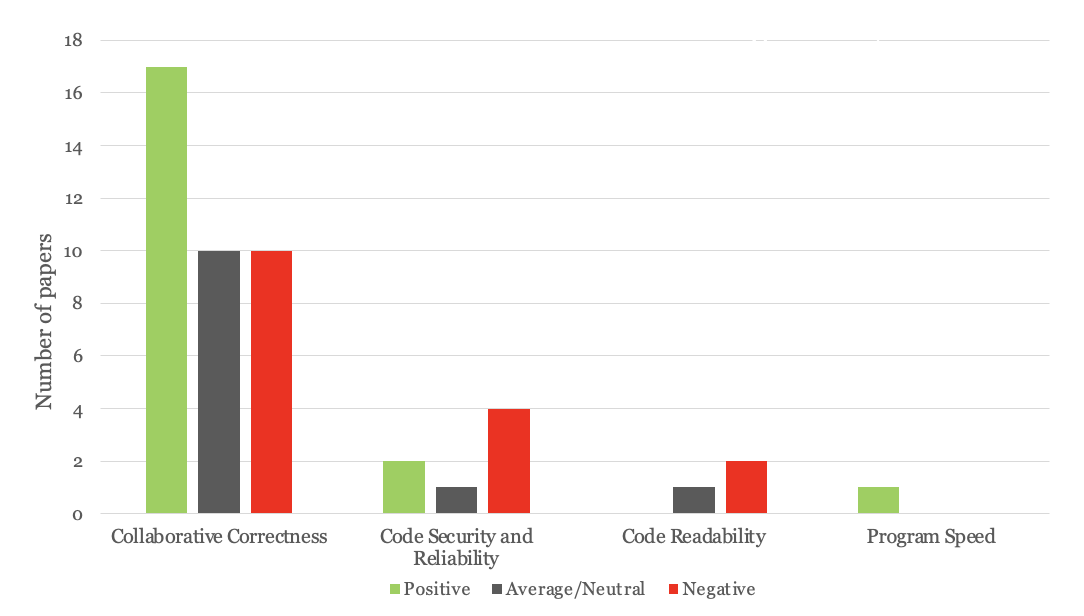}
  \caption{The LLM response Evaluation metric results as examined in the papers and grouped them into number of papers reporting positive effects, negative and neutral/average effects}
  \label{fig:response_eval_results}
\end{figure*}

\textbf{Collaborative Correctness:} 
This metric measures how closely the submitted LLM response aligns with the expected output. This was measured by rubrics, unit tests, accuracy of response, etc. We discovered a lot of variance in the accuracy of the final response. There were significant gains observed when LLMs were used for a wide variety of tasks from basic programming tasks \cite{Kazemitabaar2023ACM_CHI, Erhabor2023arXiv, Matthieu2023ACM_SPLC, Sun2024IJETHE, Nejjar2024ICSSP, Yiming2023ACM_HOTOS, Jiang2022ACM_CHI}, algorithmic tasks \cite{Liu2023ArXiv, Weisz2022ACM_IUI, Cipriano2023ACM_ITiCSE}, UML modeling \cite{Liang2024arXiv}, software engineering \cite{Acher2023SPLC, Matthieu2023ACM_SPLC, Aillon2023IEEE_C3} and Data analysis \cite{Nejjar2024ICSSP}. Even more gains were observed for specialized tools and frameworks that helped users with reviewing the LLM response (more specifically code review) \cite{Ferdowsi2023arXiv, Englhardt2023arXiv, Yan2024ACM_CHI, Kazemitabaar2024ACM_CHI}. Moreover, there were instances where the performance of the LLM was average \cite{Xue2024ACM_ICSE, Johnson2024JRTC, Wermelinger2023ACM_SIGCSE, Peng2023arXiv, Vasiliniuc2023IEEE_ICCP, Choudhuri2024ACM_ICSE, Shoufan2023ACM_TCE, Qian2024ACM_IUI, Chatterjee2024SEC, Wang2024arXiv} and instances where performance was really bad \cite{Matthieu2023ACM_SPLC, Cipriano2023ACM_ITiCSE, Zhou2023ACM_CHI_EA, Nguyen2024ACM_CHI, Nejjar2024ICSSP, Sanger2024arXiv, Perry2023ACM_CCS, Camara2023SSM, Imai2022ACM_IEEE, Aillon2023IEEE_C3}. What we noticed was that situations where either the task evaluated was very specific or the rubric was not as rigorous, there were a lot more fluctuations in the measured collaborative correctness. In studies where the measures of correctness were rigorous, accessing a variety of things or using a variety of methods, this metric normalized towards being average. 

\textbf{Code Security and Reliability:}
This metric is an evaluation of how secure and reliable the code is under a variety of conditions. It was measured via bugs detected, libraries been used, as well as code smells. 
The evaluation of this metric from the papers seems to indicate mixed results on the effect of LLMs on code security. A few studies suggest that using LLMs resulted in more secure code being generated \cite{Sandoval2023USENIX, Chatterjee2024SEC}, while others suggests that the use of LLMs resulted in more insecure code \cite{Perry2023ACM_CCS, Nejjar2024ICSSP, Oh2023IEEE_SP, Fakhoury2024arXiv}. Still another study suggests that the impact of LLMs on code security are not significant \cite{Asare2024ACM_ICSE}.
Despite these results, it is important to recognize that the number of studies where code security and reliability is evaluated explicitly are very few so there is a need for more research into this metric.

\textbf{Readability:}
This is a measure of how understandable the code is. Using halstead readability metrics, one study did not find any major difference between LLM generated code and human generated code \cite{Madi2023ACM_ASE}. Another study that looked at this metric suggests though that the readability of a piece of code seems to vary based on the task \cite{Nejjar2024ICSSP}. There is also some indication that in the code generated by the LLM, there is the possibility of the LLM generating code fragments using concepts unknown to the user \cite{Kazemitabaar2023ACM_Koli} thus rendering some parts of the code unreadable to the user without further prompting.

\textbf{Program Speed:}
This is a measure of how fast the code written with LLM runs. Only one study was found that evaluated this metric and the results of LLM's effects on runtime suggests that LLMs can improve program speed \cite{Erhabor2023arXiv}.

\begin{tcolorbox}[colback=gray!10]
\textbf{Summary of RQ3}\\
Our synthesis revealed that the effects of LLMs on task performance tends to vary. This is partly due to the diverse nature of tasks evaluated in user studies. More research is needed into this to understand the impact of LLMs on the task performance.
\end{tcolorbox}

\section{RQ4: User-LLM Interaction Effects on Human Enhancement and Task Performance}
\label{RQ4_Section}
\subsubsection{How we addressed it}
We address this by looking into studies that reported on human enhancement and task performance and check if any of those studies link an interaction behaviour to positive gains (or losses) in performance.

\subsubsection{The Results}
We found some few points indicating the effect of certain interaction behaviours on task performance. 

\subsection{Interaction Effects on Human Enhancement Evaluation}
In the analysis of interaction patterns, certain studies observed how specific interaction patterns influenced performance. For instance, users in learning/exploration mode exhibited a learning behavior as expected while using LLMs, although their time productivity decreased. Conversely, users in the implementation mode demonstrated improved time productivity (in some cases). 

\subsection{Interaction Effects on Task Performance}
With one re-prompting pattern (elaboration), we found instances where more details yielded positive results \cite{Denny2024iTCSE}.
A major interaction behaviour that reflects more positively on the prompting experience is when users are able to identify the errors correctly and also detail the method of rectifying the error \cite{Yan2024arXiv, Yin2024arXiv, Kazemitabaar2023ACM_Koli}.

While these findings shed light on the impact of interaction patterns on performance, they remain qualitative and lack methodological rigor. Notably, there is a gap in research regarding the influence of prompting patterns on performance. Further investigation is needed to comprehensively understand how interaction patterns affect performance.

\begin{tcolorbox}[colback=gray!10]
\textbf{Summary of RQ4:}\\
Studies suggest that interaction patterns can influence performance, with learning behaviors decreasing time productivity and implementation behaviors improving it. Additionally, detailed re-prompting and error correction positively affect task performance, but further research is needed to quantify these effects.
\end{tcolorbox}

\section{Discussion}
Despite the mixed results addressed in this paper, LLMs have been shown to be a powerful tool that aids in programming tasks. In this section, we discuss some of our overall observations as well as opportunities for future research.
\subsection{A Standard Set of Metrics for Evaluating User-LLM Interactions and Effects}
From our analysis in this survey paper, we observed that several studies employed different methods to measure the same metric. For example, time productivity was measured by the number of tasks completed in some studies and by completion time in others. While testing LLMs with users is still relatively new, it would be beneficial to establish a standard set of methods for objectively evaluating each metric. Researchers can then select the method that best fits their study goals. Based on the findings of this paper, we propose a set of theme metrics along with various evaluation methods for each metric. We also include recommendations on the study design for each method, specifying whether it is suitable for within-subject (WS), between-subject (BS), or both types of designs (see Table \ref{tab:suggested_human_performance}). We refrain from suggesting task evaluation metrics as these are heavily task-dependent. There are already comprehensive standard metrics available for evaluating code quality (e.g., IEEE software quality metric methodology \cite{ieee1061, IEEE730}, CERT coding standards \cite{cert_secure_coding}, etc) and we encourage researchers to use this when assessing task performance (as it relates to code). 

\begin{table*}[ht]
    \centering
    \begin{tabular}{p{3cm}p{4cm}p{4cm}p{3cm}}
        \toprule
        \textbf{Metric Theme} & \textbf{Method of evaluation} & \textbf{Definition} & \textbf{Recommended Study Design}\\
        \midrule
        \multirow{5}{*}{Time Productivity} 
                    & Success Rate & Number of tasks completed & WS, BS  \\
                    & Completion Time & Amount of time taken to complete the task & BS\\
                    & Response Length & Lines of code or number of words in user response & BS\\

        \midrule
        \multirow{4}{*}{User Frustration}
                    & Number of Errors Encountered & A count on the number of instances where users had to deal with an unexpected response from the LLM & WS, BS\\
        \midrule
        \multirow{13}{*}{Learning Check} 
                    & Pre-test and Post-test Scores & Test scores taken before (pre-test) and after (post-test) LLM use to assess its impact. The difference in these scores indicates the effect of the LLM. & BS \\
                    & Pre-test and Post-test Completion Time & Completion time taken before (pre-test) and after (post-test) LLM use to assess its impact. The difference in these scores indicates the effect of the LLM. & BS  \\
                    & Computational thinking measure (Pre and Post tests) \cite{Korkmaz2017CHB} & Computational thinking score difference before (pre-test) and after (post-test) LLM use to assess its impact.  & BS  \\
        \bottomrule
    \end{tabular}
    \caption{Suggested human enhancement metric themes along with recommended methods of evaluation. BS means between subjects study design and WS means within subjects study design}
    \label{tab:suggested_human_performance}
\end{table*}

\subsection{Managing the Non-determinism of LLMs}
One of the things we know definitively about LLMs is that they are non-deterministic in nature. Thus, figuring out how to interact with LLMs to get the desired response is incredibly valuable. 

We've shifted from learning program syntax to learning how to communicate code in natural language, which should come easier since we use natural language daily. However, research shows that users of LLMs struggle with this. It's crucial for humans to understand effective interaction techniques with LLMs to produce the needed response. The section on re-prompting strategies in our response to RQ1 is beneficial for this purpose, serving as a starting point for daily users when interacting with LLMs. 

Our analysis in response to RQ1 reveals a paradigm consisting of three considerations when interacting with LLMs:
\begin{enumerate}
    \item \textbf{What is the problem you want the LLM to solve?} This directly determines the type of request made to the LLM; whether it is a learning request, implementation request, or error-correcting request.
    \item \textbf{How to input the request (or prompt) into the LLM:} Users should choose a prompting style (single or multi-prompt) and a prompting strategy (or strategies) as outlined in \ref{RQ1_Section} as a starting point for prompting or as methods to re-prompt if the LLM fails to produce a correct response. White et al. has also listed out some prompting patterns that can be used as well when trying to get a desired response from the LLM \cite{white2023arXiv}.
    \item In cases where the LLM continues to produce incorrect responses after multiple attempts, users may need consider other resources. The LLM may not have a "correct" answer, or the user's background knowledge on the problem might need improvement to achieve better results.
\end{enumerate}

Starting with these key considerations can enhance the effectiveness of LLM usage. By adopting effective re-prompting strategies and being aware of the nuances in interacting with LLMs, users can better leverage these powerful tools to achieve their desired outcomes. Future research should continue to explore and refine these interaction techniques to maximize the potential of LLMs in various applications.

\subsection{LLMs for Learning}
\label{sec:llms-for-learning}
The research suggests that LLMs can play a role in facilitating learning about programming techniques and concepts in certain contexts. This literature survey has shown that programmers need not excessively worry about impeding their learning progress, though results are mixed and the extent of learning benefit appears to depend considerably on how LLMs are implemented and used. However, some of the concerns regarding LLMs often revolve around academic integrity and the potential for plagiarism, prompting educators to seek alternative assessment methods that mitigate these concerns. Presently, educational policies often entail educators prohibiting the utilization of LLMs for assignments. It's essential for students to adhere to existing institutional policies governing LLM usage, although it's worth noting that these policies may evolve over time to reflect a better understanding of LLMs. One approach for educators to address learning concerns is to blend in-class assessments (without LLM usage) with homework assignments where students are permitted to freely utilize LLMs to enhance their understanding of course concepts.
Additionally, it is important that learning goals (whether for a course or for self-study) are set properly. There is a difference between using LLMs to learn how to program and using LLMs to understand programming. Users (and educators) need to be cautious of this as the use of LLMs will definitely help students in understanding programs but not necessarily in learning how to program. This distinction is further reflected in our Learning Check findings — the studies that reported more consistent learning gains tended to involve structured or guided LLM implementations, such as purpose-built tutoring tools, rather than unstructured access to a general-purpose LLM. Future research should investigate implementation structure as a moderating variable for learning outcomes.

\subsection{Opportunities for Future Work for Researchers}
\subsubsection{Investigating the Impact of Interaction Patterns on Performance:}The insights from RQ1 reveal current user-LLM interactions and highlight challenges programmers face. These findings are useful for LLM developers to address these issues. However, a full evaluation of how these interactions impact performance is still missing. Future research can explore the interaction behaviours that influence task performance and human enhancement metrics. Additionally, some studies outside the scope of this paper have discussed LLM interaction patterns, but their effectiveness across different tasks has not been tested \cite{white2023arXiv}.

\subsubsection{Validating Findings Across LLM Models:}
Additionally, it's essential to acknowledge that certain types of LLMs, such as T5, StarCoder, have not undergone user studies (or at least none were found in our literature survey). Most of the LLMs used in this study were based off ChatGPT or copilot. This underscores the necessity for future research to validate the generalizability of findings across different LLM models. This gap in knowledge presents an opportunity for further exploration to ensure robust insights into LLM usability and performance.

\section{Related Work}
Some literature reviews on LLMs have predominantly addressed the concerns of the machine learning community, focusing on factors such as model size, data quality, and expert tuning, as emphasized by Chen et al. \cite{chen2023large}. However, these insights often lack actionable guidance for programmers seeking to integrate LLMs into their workflow efficiently. Similarly, Wong et al. \cite{Wong2023Entropy} explored downstream tasks enabled by LLMs but lacked a user-centered perspective.

While Zhang et al. \cite{Zhang2023arXiv} extensively covered LLM applications in software engineering, including tasks examined with LLMs, our study takes a distinct approach by delving into user interaction and LLM usability. We specifically investigate task completion abilities and interaction patterns, providing insights into user-centric usability scenarios that complement existing research. Similarly, Yang et al. \cite{Yang2024ArXiv} explored LLM performance metrics and usability studies but did not delve as deeply into non-AI expert usability scenarios as our investigation.

Conversely, Sarkar et al. \cite{Sarkar2022arXiv} concentrated on LLM usability for programming tasks, emphasizing user studies and experience reports with a qualitative approach. In contrast, our study employs both qualitative and quantitative methodologies to comprehensively understand how users currently engage with LLMs and identify opportunities for integration into their workflows to enhance task performance.

\section{Threats to Validity}
\subsection{Search Methodology} During our literature survey, we had to choose search terms carefully to find papers on human usability studies. It was tough balancing inclusivity with relevance, as we wanted to avoid getting too many irrelevant papers. However, we managed this challenge by carefully screening search results based on our study criteria. This helped us find valuable insights while keeping our review on track. We also employed backward snowballing as well as forward snowballing to account for missed search terms. The diversity in the types of papers we found is proof that the strategy is valid.

\subsection{Qualitative Data Analysis} Aggregating qualitative data from multiple papers presents another potential threat to validity. While there is a risk of overlooking insights, we have addressed this by clearly indicating any patterns with limited evidence and ensuring that omitted insights do not compromise the validity of our results.

\section{Conclusion}
In this paper, we conducted a literature survey to examine user studies that assess the interactions between humans and LLMs as well as identify human enhancements and task performance effects as a result of this interaction. Our study addressed four key research questions: the nature of programmers' interactions with LLMs, the enhancement of human capabilities through LLM use, the improvement in task performance due to LLMs, and the interaction behaviors that lead to these enhancements and improvements.

Our analysis revealed diverse objectives and methodologies in existing studies, highlighting the need for standardized metrics to evaluate user-LLM interactions objectively. We identified common human enhancement metrics such as time productivity and learning outcomes and task performance metrics like code quality and correctness. The non-deterministic nature of LLMs necessitates effective interaction techniques, including re-prompting strategies, which as we discussed, are crucial for users to achieve desired responses.

The research also underscored the role of LLMs in facilitating learning about programming techniques and concepts, though concerns about academic integrity remain. Future research opportunities include investigating the impact of specific interaction patterns on performance and validating findings across different LLM models.

In summary, our survey provides valuable insights into user-LLM interactions and their effects on human and task performance. Importantly, our findings highlight that the effects of LLMs on productivity and learning are context-dependent and mixed, underscoring the need for careful interpretation of results in this space and cautioning against overgeneralizing positive outcomes. By establishing standard evaluation methods and refining interaction techniques, we can better leverage LLMs' potential in various applications and enhance our understanding of their capabilities and limitations.

\begin{acks}
The authors would like to thank Peter Mawhorter for his careful reading of this work and for bringing some inaccuracies to our attention. His feedback led to meaningful improvements in the accuracy and clarity of our findings.
\end{acks}

\bibliographystyle{ACM-Reference-Format}
\bibliography{main}

@String{Computing = "Computing" }

@String{Computer = "{IEEE} Computer" }

@inproceedings{Ozturk2023EICC,
author = {Ozturk, Omer Said and Ekmekcioglu, Emre and Cetin, Orcun and Arief, Budi and Hernandez-Castro, Julio},
title = {New Tricks to Old Codes: Can AI Chatbots Replace Static Code Analysis Tools?},
year = {2023},
isbn = {9781450398299},
publisher = {Association for Computing Machinery},
address = {New York, NY, USA},
url = {https://doi.org/10.1145/3590777.3590780},
doi = {10.1145/3590777.3590780},
booktitle = {Proceedings of the 2023 European Interdisciplinary Cybersecurity Conference},
pages = {13–18},
numpages = {6},
location = {Stavanger, Norway},
series = {EICC '23}
}

@inproceedings{Byun2023ACM_CHI,
author = {Byun, Courtni and Vasicek, Piper and Seppi, Kevin},
title = {Dispensing with Humans in Human-Computer Interaction Research},
year = {2023},
isbn = {9781450394222},
publisher = {Association for Computing Machinery},
address = {New York, NY, USA},
url = {https://doi.org/10.1145/3544549.3582749},
doi = {10.1145/3544549.3582749},
booktitle = {Extended Abstracts of the 2023 CHI Conference on Human Factors in Computing Systems},
articleno = {413},
numpages = {26},
location = {Hamburg, Germany},
series = {CHI EA '23}
}

@article{Tian2023ArXiv,
Author = {Haoye Tian and Weiqi Lu and Tsz On Li and Xunzhu Tang and Shing-Chi Cheung and Jacques Klein and Tegawendé F. Bissyandé},
Title = {Is ChatGPT the Ultimate Programming Assistant -- How far is it?},
Year = {2023},
Eprint = {https://arxiv.org/abs/2304.11938},
primaryClass={cs.SE},
}

@inproceedings{Savelka2023ACM_ITiCSE,
author = {Savelka, Jaromir and Agarwal, Arav and Bogart, Christopher and Song, Yifan and Sakr, Majd},
title = {Can Generative Pre-Trained Transformers (GPT) Pass Assessments in Higher Education Programming Courses?},
year = {2023},
isbn = {9798400701382},
publisher = {Association for Computing Machinery},
address = {New York, NY, USA},
url = {https://doi.org/10.1145/3587102.3588792},
doi = {10.1145/3587102.3588792},
booktitle = {Proceedings of the 2023 Conference on Innovation and Technology in Computer Science Education V. 1},
pages = {117–123},
numpages = {7},
location = {Turku, Finland},
series = {ITiCSE 2023}
}

@inproceedings{Ouh2023ACM_ITiCSE,
author = {Ouh, Eng Lieh and Gan, Benjamin Kok Siew and Jin Shim, Kyong and Wlodkowski, Swavek},
title = {ChatGPT, Can You Generate Solutions for My Coding Exercises? An Evaluation on Its Effectiveness in an Undergraduate Java Programming Course.},
year = {2023},
isbn = {9798400701382},
publisher = {Association for Computing Machinery},
address = {New York, NY, USA},
url = {https://doi.org/10.1145/3587102.3588794},
doi = {10.1145/3587102.3588794},
booktitle = {Proceedings of the 2023 Conference on Innovation and Technology in Computer Science Education V. 1},
pages = {54–60},
numpages = {7},
location = {Turku, Finland},
series = {ITiCSE 2023}
}

@inproceedings{FinnieAinsley2023ACM_ACE,
author = {Finnie-Ansley, James and Denny, Paul and Becker, Brett A. and Luxton-Reilly, Andrew and Prather, James},
title = {The Robots Are Coming: Exploring the Implications of OpenAI Codex on Introductory Programming},
year = {2022},
isbn = {9781450396431},
publisher = {Association for Computing Machinery},
address = {New York, NY, USA},
url = {https://doi.org/10.1145/3511861.3511863},
doi = {10.1145/3511861.3511863},
booktitle = {Proceedings of the 24th Australasian Computing Education Conference},
pages = {10–19},
numpages = {10},
location = {Virtual Event, Australia},
series = {ACE '22}
}

@misc{Ma2023arXiv,
      title={HypoCompass: Large-Language-Model-based Tutor for Hypothesis Construction in Debugging for Novices}, 
      author={Qianou Ma and Hua Shen and Kenneth Koedinger and Tongshuang Wu},
      year={2023},
      eprint={2310.05292},
      archivePrefix={arXiv},
      primaryClass={cs.HC}
}

@inproceedings{Zamfirescu2023ACM_CHI,
author = {Zamfirescu-Pereira, J.D. and Wong, Richmond Y. and Hartmann, Bjoern and Yang, Qian},
title = {Why Johnny Can’t Prompt: How Non-AI Experts Try (and Fail) to Design LLM Prompts},
year = {2023},
isbn = {9781450394215},
publisher = {Association for Computing Machinery},
address = {New York, NY, USA},
url = {https://doi.org/10.1145/3544548.3581388},
doi = {10.1145/3544548.3581388},
booktitle = {Proceedings of the 2023 CHI Conference on Human Factors in Computing Systems},
articleno = {437},
numpages = {21},
location = {Hamburg, Germany},
series = {CHI '23}
}

@article{Barke2023ACM_PL,
author = {Barke, Shraddha and James, Michael B. and Polikarpova, Nadia},
title = {Grounded Copilot: How Programmers Interact with Code-Generating Models},
year = {2023},
issue_date = {April 2023},
publisher = {Association for Computing Machinery},
address = {New York, NY, USA},
volume = {7},
number = {OOPSLA1},
url = {https://doi.org/10.1145/3586030},
doi = {10.1145/3586030},
journal = {Proc. ACM Program. Lang.},
month = {apr},
articleno = {78},
numpages = {27}
}

@misc{Kazemitabaar2023ACM_Koli,
      title={How Novices Use LLM-Based Code Generators to Solve CS1 Coding Tasks in a Self-Paced Learning Environment}, 
      author={Majeed Kazemitabaar and Xinying Hou and Austin Henley and Barbara J. Ericson and David Weintrop and Tovi Grossman},
      year={2023},
      eprint={2309.14049},
      archivePrefix={arXiv},
      primaryClass={cs.HC}
}

@inproceedings{Vaithilingam2022ACM_CHI_EA,
author = {Vaithilingam, Priyan and Zhang, Tianyi and Glassman, Elena L.},
title = {Expectation vs. Experience: Evaluating the Usability of Code Generation Tools Powered by Large Language Models},
year = {2022},
isbn = {9781450391566},
publisher = {Association for Computing Machinery},
address = {New York, NY, USA},
url = {https://doi.org/10.1145/3491101.3519665},
doi = {10.1145/3491101.3519665},
booktitle = {Extended Abstracts of the 2022 CHI Conference on Human Factors in Computing Systems},
articleno = {332},
numpages = {7},
location = {New Orleans, LA, USA},
series = {CHI EA '22}
}

@inproceedings{Kazemitabaar2023ACM_CHI,
author = {Kazemitabaar, Majeed and Chow, Justin and Ma, Carl Ka To and Ericson, Barbara J. and Weintrop, David and Grossman, Tovi},
title = {Studying the Effect of AI Code Generators on Supporting Novice Learners in Introductory Programming},
year = {2023},
isbn = {9781450394215},
publisher = {Association for Computing Machinery},
address = {New York, NY, USA},
url = {https://doi.org/10.1145/3544548.3580919},
doi = {10.1145/3544548.3580919},
booktitle = {Proceedings of the 2023 CHI Conference on Human Factors in Computing Systems},
articleno = {455},
numpages = {23},
location = {Hamburg, Germany},
series = {CHI '23}
}

@article{Arghavan2023JSS,
title = {GitHub Copilot AI pair programmer: Asset or Liability?},
journal = {Journal of Systems and Software},
volume = {203},
pages = {111734},
year = {2023},
issn = {0164-1212},
doi = {https://doi.org/10.1016/j.jss.2023.111734},
url = {https://www.sciencedirect.com/science/article/pii/S0164121223001292},
author = {Arghavan {Moradi Dakhel} and Vahid Majdinasab and Amin Nikanjam and Foutse Khomh and Michel C. Desmarais and Zhen Ming (Jack) Jiang}
}

@article{Shoufan2023ACM_TCE,
author = {Shoufan, Abdulhadi},
title = {Can Students without Prior Knowledge Use ChatGPT to Answer Test Questions? An Empirical Study},
year = {2023},
publisher = {Association for Computing Machinery},
address = {New York, NY, USA},
url = {https://doi.org/10.1145/3628162},
doi = {10.1145/3628162},
note = {Just Accepted},
journal = {ACM Trans. Comput. Educ.},
month = {oct}
}

@article{Prather2023ACM_TCI,
author = {Prather, James and Reeves, Brent N. and Denny, Paul and Becker, Brett A. and Leinonen, Juho and Luxton-Reilly, Andrew and Powell, Garrett and Finnie-Ansley, James and Santos, Eddie Antonio},
title = {“It’s Weird That It Knows What I Want”: Usability and Interactions with Copilot for Novice Programmers},
year = {2023},
publisher = {Association for Computing Machinery},
address = {New York, NY, USA},
issn = {1073-0516},
url = {https://doi.org/10.1145/3617367},
doi = {10.1145/3617367},
note = {Just Accepted},
journal = {ACM Trans. Comput.-Hum. Interact.},
month = {aug}
}

@inproceedings{Madi2023ACM_ASE,
author = {Al Madi, Naser},
title = {How Readable is Model-Generated Code? Examining Readability and Visual Inspection of GitHub Copilot},
year = {2023},
isbn = {9781450394758},
publisher = {Association for Computing Machinery},
address = {New York, NY, USA},
url = {https://doi.org/10.1145/3551349.3560438},
doi = {10.1145/3551349.3560438},
booktitle = {Proceedings of the 37th IEEE/ACM International Conference on Automated Software Engineering},
articleno = {205},
numpages = {5},
location = {Rochester, MI, USA},
series = {ASE '22}
}

@inproceedings{Jayagopal2022ACM_UIST,
author = {Jayagopal, Dhanya and Lubin, Justin and Chasins, Sarah E.},
title = {Exploring the Learnability of Program Synthesizers by Novice Programmers},
year = {2022},
isbn = {9781450393201},
publisher = {Association for Computing Machinery},
address = {New York, NY, USA},
url = {https://doi.org/10.1145/3526113.3545659},
doi = {10.1145/3526113.3545659},
booktitle = {Proceedings of the 35th Annual ACM Symposium on User Interface Software and Technology},
articleno = {64},
numpages = {15},
location = {Bend, OR, USA},
series = {UIST '22}
}

@inproceedings{Weisz2022ACM_IUI,
author = {Weisz, Justin D. and Muller, Michael and Ross, Steven I. and Martinez, Fernando and Houde, Stephanie and Agarwal, Mayank and Talamadupula, Kartik and Richards, John T.},
title = {Better Together? An Evaluation of AI-Supported Code Translation},
year = {2022},
isbn = {9781450391443},
publisher = {Association for Computing Machinery},
address = {New York, NY, USA},
url = {https://doi.org/10.1145/3490099.3511157},
doi = {10.1145/3490099.3511157},
booktitle = {27th International Conference on Intelligent User Interfaces},
pages = {369–391},
numpages = {23},
location = {Helsinki, Finland},
series = {IUI '22}
}

@inproceedings{chen2023large,
author = {Chen, Bei and Zan, Daoguang and Zhang, Fengji and Lu, Dianjie and Wu, Bingchao and Guan, Bei and Wang, Yongji and Lou, Jian-Guang},
title = {Large Language Models Meet NL2Code: A Survey},
booktitle = {ACL 2023},
year = {2023},
month = {June},
url = {https://www.microsoft.com/en-us/research/publication/large-language-models-meet-nl2code-a-survey/},
}

@INPROCEEDINGS{Imai2022ACM_IEEE,
  author={Imai, Saki},
  booktitle={2022 IEEE/ACM 44th International Conference on Software Engineering: Companion Proceedings (ICSE-Companion)}, 
  title={Is GitHub Copilot a Substitute for Human Pair-programming? An Empirical Study}, 
  year={2022},
  volume={},
  number={},
  pages={319-321},
  doi={10.1145/3510454.3522684}
}

@inproceedings {Sandoval2023USENIX,
	author = {Gustavo Sandoval and Hammond Pearce and Teo Nys and Ramesh Karri and Siddharth Garg and Brendan Dolan-Gavitt},
	title = {Lost at C: A User Study on the Security Implications of Large Language Model Code Assistants},
	booktitle = {32nd USENIX Security Symposium (USENIX Security 23)},
	year = {2023},
	isbn = {978-1-939133-37-3},
	address = {Anaheim, CA},
	pages = {2205--2222},
	url = {https://www.usenix.org/conference/usenixsecurity23/presentation/sandoval},
	publisher = {USENIX Association},
	month = aug
}

@misc{Englhardt2023arXiv,
      title={Exploring and Characterizing Large Language Models For Embedded System Development and Debugging}, 
      author={Zachary Englhardt and Richard Li and Dilini Nissanka and Zhihan Zhang and Girish Narayanswamy and Joseph Breda and Xin Liu and Shwetak Patel and Vikram Iyer},
      year={2023},
      eprint={2307.03817},
      archivePrefix={arXiv},
      primaryClass={cs.SE}
}

@misc{Yen2023arXiv,
      title={CoLadder: Supporting Programmers with Hierarchical Code Generation in Multi-Level Abstraction}, 
      author={Ryan Yen and Jiawen Zhu and Sangho Suh and Haijun Xia and Jian Zhao},
      year={2023},
      eprint={2310.08699},
      archivePrefix={arXiv},
      primaryClass={cs.SE}
}

@misc{Peng2023arXiv,
      title={The Impact of AI on Developer Productivity: Evidence from GitHub Copilot}, 
      author={Sida Peng and Eirini Kalliamvakou and Peter Cihon and Mert Demirer},
      year={2023},
      eprint={2302.06590},
      archivePrefix={arXiv},
      primaryClass={cs.SE}
}

@inproceedings{Kim2022ACM_CHI,
author = {Kim, Tae Soo and Choi, DaEun and Choi, Yoonseo and Kim, Juho},
title = {Stylette: Styling the Web with Natural Language},
year = {2022},
isbn = {9781450391573},
publisher = {Association for Computing Machinery},
address = {New York, NY, USA},
url = {https://doi.org/10.1145/3491102.3501931},
doi = {10.1145/3491102.3501931},
booktitle = {Proceedings of the 2022 CHI Conference on Human Factors in Computing Systems},
articleno = {5},
numpages = {17},
location = {, New Orleans, LA, USA, },
series = {CHI '22}
}

@article{Bird2023ACM_Queue,
author = {Bird, Christian and Ford, Denae and Zimmermann, Thomas and Forsgren, Nicole and Kalliamvakou, Eirini and Lowdermilk, Travis and Gazit, Idan},
title = {Taking Flight with Copilot: Early insights and opportunities of AI-powered pair-programming tools},
year = {2023},
issue_date = {November/December},
publisher = {Association for Computing Machinery},
address = {New York, NY, USA},
volume = {20},
number = {6},
issn = {1542-7730},
url = {https://doi.org/10.1145/3582083},
doi = {10.1145/3582083},
journal = {Queue},
month = {jan},
pages = {35–57},
numpages = {23}
}

@inproceedings{Matthieu2023ACM_SPLC,
author = {Acher, Mathieu and Martinez, Jabier},
title = {Generative AI for Reengineering Variants into Software Product Lines: An Experience Report},
year = {2023},
isbn = {9798400700927},
publisher = {Association for Computing Machinery},
address = {New York, NY, USA},
url = {https://doi.org/10.1145/3579028.3609016},
doi = {10.1145/3579028.3609016},
booktitle = {Proceedings of the 27th ACM International Systems and Software Product Line Conference - Volume B},
pages = {57–66},
numpages = {10},
location = {Tokyo, Japan},
series = {SPLC '23}
}

@inproceedings{Cipriano2023ACM_ITiCSE,
author = {Cipriano, Bruno Pereira and Alves, Pedro},
title = {GPT-3 vs Object Oriented Programming Assignments: An Experience Report},
year = {2023},
isbn = {9798400701382},
publisher = {Association for Computing Machinery},
address = {New York, NY, USA},
url = {https://doi.org/10.1145/3587102.3588814},
doi = {10.1145/3587102.3588814},
booktitle = {Proceedings of the 2023 Conference on Innovation and Technology in Computer Science Education V. 1},
pages = {61–67},
numpages = {7},
location = {, Turku, Finland, },
series = {ITiCSE 2023}
}

@inproceedings{Yiming2023ACM_HOTOS,
author = {Su, Yiming and Wan, Chengcheng and Sethi, Utsav and Lu, Shan and Musuvathi, Madan and Nath, Suman},
title = {HotGPT: How to Make Software Documentation More Useful with a Large Language Model?},
year = {2023},
isbn = {9798400701955},
publisher = {Association for Computing Machinery},
address = {New York, NY, USA},
url = {https://doi.org/10.1145/3593856.3595910},
doi = {10.1145/3593856.3595910},
booktitle = {Proceedings of the 19th Workshop on Hot Topics in Operating Systems},
pages = {87–93},
numpages = {7},
location = {Providence, RI, USA},
series = {HOTOS '23}
}

@inproceedings{Liu2023ACM_CHI,
author = {Liu, Michael Xieyang and Sarkar, Advait and Negreanu, Carina and Zorn, Benjamin and Williams, Jack and Toronto, Neil and Gordon, Andrew D.},
title = {“What It Wants Me To Say”: Bridging the Abstraction Gap Between End-User Programmers and Code-Generating Large Language Models},
year = {2023},
isbn = {9781450394215},
publisher = {Association for Computing Machinery},
address = {New York, NY, USA},
url = {https://doi.org/10.1145/3544548.3580817},
doi = {10.1145/3544548.3580817},
booktitle = {Proceedings of the 2023 CHI Conference on Human Factors in Computing Systems},
articleno = {598},
numpages = {31},
location = {, Hamburg, Germany, },
series = {CHI '23}
}

@inproceedings{Zhou2023ACM_CHI_EA,
author = {Zhou, Haoquan and Li, Jingbo},
title = {A Case Study on Scaffolding Exploratory Data Analysis for AI Pair Programmers},
year = {2023},
isbn = {9781450394222},
publisher = {Association for Computing Machinery},
address = {New York, NY, USA},
url = {https://doi.org/10.1145/3544549.3583943},
doi = {10.1145/3544549.3583943},
booktitle = {Extended Abstracts of the 2023 CHI Conference on Human Factors in Computing Systems},
articleno = {561},
numpages = {7},
location = {, Hamburg, Germany, },
series = {CHI EA '23}
}

@inproceedings{Ross2023ACM_IUI,
author = {Ross, Steven I. and Martinez, Fernando and Houde, Stephanie and Muller, Michael and Weisz, Justin D.},
title = {The Programmer’s Assistant: Conversational Interaction with a Large Language Model for Software Development},
year = {2023},
isbn = {9798400701061},
publisher = {Association for Computing Machinery},
address = {New York, NY, USA},
url = {https://doi.org/10.1145/3581641.3584037},
doi = {10.1145/3581641.3584037},
booktitle = {Proceedings of the 28th International Conference on Intelligent User Interfaces},
pages = {491–514},
numpages = {24},
location = {Sydney, NSW, Australia},
series = {IUI '23}
}

@article{Yilmaz2023CEAI ,
title = {The effect of generative artificial intelligence (AI)-based tool use on students' computational thinking skills, programming self-efficacy and motivation},
journal = {Computers and Education: Artificial Intelligence},
volume = {4},
pages = {100147},
year = {2023},
issn = {2666-920X},
doi = {https://doi.org/10.1016/j.caeai.2023.100147},
url = {https://www.sciencedirect.com/science/article/pii/S2666920X23000267},
author = {Ramazan Yilmaz and Fatma Gizem {Karaoglan Yilmaz}}
}

@inproceedings{MacNeil2023ACM_SIGCSE,
author = {MacNeil, Stephen and Tran, Andrew and Hellas, Arto and Kim, Joanne and Sarsa, Sami and Denny, Paul and Bernstein, Seth and Leinonen, Juho},
title = {Experiences from Using Code Explanations Generated by Large Language Models in a Web Software Development E-Book},
year = {2023},
isbn = {9781450394314},
publisher = {Association for Computing Machinery},
address = {New York, NY, USA},
url = {https://doi.org/10.1145/3545945.3569785},
doi = {10.1145/3545945.3569785},
booktitle = {Proceedings of the 54th ACM Technical Symposium on Computer Science Education V. 1},
pages = {931–937},
numpages = {7},
location = {, Toronto ON, Canada, },
series = {SIGCSE 2023}
}

@misc{Feng2023arXiv,
      title={CoPrompt: Supporting Prompt Sharing and Referring in Collaborative Natural Language Programming}, 
      author={Felicia Li Feng and Ryan Yen and Yuzhe You and Mingming Fan and Jian Zhao and Zhicong Lu},
      year={2023},
      eprint={2310.09235},
      archivePrefix={arXiv},
      primaryClass={cs.HC}
}

@misc{YM2023arXiv,
      title={PwR: Exploring the Role of Representations in Conversational Programming}, 
      author={Pradyumna YM and Vinod Ganesan and Dinesh Kumar Arumugam and Meghna Gupta and Nischith Shadagopan and Tanay Dixit and Sameer Segal and Pratyush Kumar and Mohit Jain and Sriram Rajamani},
      year={2023},
      eprint={2309.09495},
      archivePrefix={arXiv},
      primaryClass={cs.HC}
}

@inproceedings{Wermelinger2023ACM_SIGCSE,
author = {Wermelinger, Michel},
title = {Using GitHub Copilot to Solve Simple Programming Problems},
year = {2023},
isbn = {9781450394314},
publisher = {Association for Computing Machinery},
address = {New York, NY, USA},
url = {https://doi.org/10.1145/3545945.3569830},
doi = {10.1145/3545945.3569830},
booktitle = {Proceedings of the 54th ACM Technical Symposium on Computer Science Education V. 1},
pages = {172–178},
numpages = {7},
location = {, Toronto ON, Canada, },
series = {SIGCSE 2023}
}

@inproceedings{Jiang2022ACM_CHI,
author = {Jiang, Ellen and Toh, Edwin and Molina, Alejandra and Olson, Kristen and Kayacik, Claire and Donsbach, Aaron and Cai, Carrie J and Terry, Michael},
title = {Discovering the Syntax and Strategies of Natural Language Programming with Generative Language Models},
year = {2022},
isbn = {9781450391573},
publisher = {Association for Computing Machinery},
address = {New York, NY, USA},
url = {https://doi.org/10.1145/3491102.3501870},
doi = {10.1145/3491102.3501870},
booktitle = {Proceedings of the 2022 CHI Conference on Human Factors in Computing Systems},
articleno = {386},
numpages = {19},
location = {, New Orleans, LA, USA, },
series = {CHI '22}
}

@inproceedings{Acher2023SPLC,
author = {Acher, Mathieu and Duarte, Jos\'{e} Galindo and J\'{e}z\'{e}quel, Jean-Marc},
title = {On Programming Variability with Large Language Model-based Assistant},
year = {2023},
isbn = {9798400700910},
publisher = {Association for Computing Machinery},
address = {New York, NY, USA},
url = {https://doi.org/10.1145/3579027.3608972},
doi = {10.1145/3579027.3608972},
booktitle = {Proceedings of the 27th ACM International Systems and Software Product Line Conference - Volume A},
pages = {8–14},
numpages = {7},
location = {Tokyo, Japan},
series = {SPLC '23}
}

@misc{Mozanner2023rXiv,
      title={Reading Between the Lines: Modeling User Behavior and Costs in AI-Assisted Programming}, 
      author={Hussein Mozannar and Gagan Bansal and Adam Fourney and Eric Horvitz},
      year={2023},
      eprint={2210.14306},
      archivePrefix={arXiv},
      primaryClass={cs.SE}
}

@article{Tang2023CMU,
author = "Ningzhi Tang and Meng Chen and Zheng Ning and Aakash Bansal and Yu Huang and Collin McMillan and Toby Jia-Jun Li",
title = "{An Empirical Study of Developer Behaviors for Validating and Repairing AI-Generated Code}",
year = "2023",
month = "3",
url = "https://kilthub.cmu.edu/articles/conference_contribution/An_Empirical_Study_of_Developer_Behaviors_for_Validating_and_Repairing_AI-Generated_Code/22223533",
doi = "10.1184/R1/22223533.v1"
}

@misc{Ferdowsi2023arXiv,
      title={Live Exploration of AI-Generated Programs}, 
      author={Kasra Ferdowsi and Ruanqianqian Huang and Michael B. James and Nadia Polikarpova and Sorin Lerner},
      year={2023},
      eprint={2306.09541},
      archivePrefix={arXiv},
      primaryClass={cs.HC}
}

@inproceedings{Prasad2023ACM_CompEd,
author = {Prasad, Siddhartha and Greenman, Ben and Nelson, Tim and Krishnamurthi, Shriram},
title = {Generating Programs Trivially: Student Use of Large Language Models},
year = {2023},
isbn = {9798400700484},
publisher = {Association for Computing Machinery},
address = {New York, NY, USA},
url = {https://doi.org/10.1145/3576882.3617921},
doi = {10.1145/3576882.3617921},
booktitle = {Proceedings of the ACM Conference on Global Computing Education Vol 1},
pages = {126–132},
numpages = {7},
location = {, Hyderabad, India, },
series = {CompEd 2023}
}

@inproceedings{Vaswani2017NIPS,
 author = {Vaswani, Ashish and Shazeer, Noam and Parmar, Niki and Uszkoreit, Jakob and Jones, Llion and Gomez, Aidan N and Kaiser, \L ukasz and Polosukhin, Illia},
 booktitle = {Advances in Neural Information Processing Systems},
 editor = {I. Guyon and U. Von Luxburg and S. Bengio and H. Wallach and R. Fergus and S. Vishwanathan and R. Garnett},
 pages = {},
 publisher = {Curran Associates, Inc.},
 title = {Attention is All you Need},
 url = {https://proceedings.neurips.cc/paper_files/paper/2017/file/3f5ee243547dee91fbd053c1c4a845aa-Paper.pdf},
 volume = {30},
 year = {2017}
}

@misc{Zhang2023arXiv,
      title={A Survey on Large Language Models for Software Engineering}, 
      author={Quanjun Zhang and Chunrong Fang and Yang Xie and Yaxin Zhang and Yun Yang and Weisong Sun and Shengcheng Yu and Zhenyu Chen},
      year={2023},
      eprint={2312.15223},
      archivePrefix={arXiv},
      primaryClass={cs.SE}
}

@misc{Sarkar2022arXiv,
      title={What is it like to program with artificial intelligence?}, 
      author={Advait Sarkar and Andrew D. Gordon and Carina Negreanu and Christian Poelitz and Sruti Srinivasa Ragavan and Ben Zorn},
      year={2022},
      eprint={2208.06213},
      archivePrefix={arXiv},
      primaryClass={cs.HC}
}

@misc{Dreyfus1980,
      author = {Dreyfus, S.E. and Dreyfus, Hubert},
year = {1980},
month = {02},
pages = {22},
title = {A Five-Stage Model of the Mental Activities Involved in Directed Skill Acquisition},
url = {https://apps.dtic.mil/sti/citations/ADA084551#:~:text=In%20acquiring%20a%20skill%20by,%2C%20proficiency%2C%20expertise%20and%20mastery.}
}

@misc{Seth2010,
      author = {Seth},
year = {2010},
month = {10},
pages = {12},
title = {Getting smart about the hierarchy of smart},
url = {https://seths.blog/2010/10/getting-smart-about-the-hierarchy-of-smart/}
}

@inproceedings{Pu2023ACM_UIST,
author = {Pu, Kevin and Yang, Jim and Yuan, Angel and Ma, Minyi and Dong, Rui and Wang, Xinyu and Chen, Yan and Grossman, Tovi},
title = {DiLogics: Creating Web Automation Programs with Diverse Logics},
year = {2023},
isbn = {9798400701320},
publisher = {Association for Computing Machinery},
address = {New York, NY, USA},
url = {https://doi.org/10.1145/3586183.3606822},
doi = {10.1145/3586183.3606822},
booktitle = {Proceedings of the 36th Annual ACM Symposium on User Interface Software and Technology},
articleno = {74},
numpages = {15},
location = {, San Francisco, CA, USA, },
series = {UIST '23}
}

@inproceedings{Xia2023ACM_ICSE,
author = {Xia, Chunqiu Steven and Wei, Yuxiang and Zhang, Lingming},
title = {Automated Program Repair in the Era of Large Pre-Trained Language Models},
year = {2023},
isbn = {9781665457019},
publisher = {IEEE Press},
url = {https://doi.org/10.1109/ICSE48619.2023.00129},
doi = {10.1109/ICSE48619.2023.00129},
booktitle = {Proceedings of the 45th International Conference on Software Engineering},
pages = {1482–1494},
numpages = {13},
location = {Melbourne, Victoria, Australia},
series = {ICSE '23}
}

@Article{Wong2023Entropy,
AUTHOR = {Wong, Man-Fai and Guo, Shangxin and Hang, Ching-Nam and Ho, Siu-Wai and Tan, Chee-Wei},
TITLE = {Natural Language Generation and Understanding of Big Code for AI-Assisted Programming: A Review},
JOURNAL = {Entropy},
VOLUME = {25},
YEAR = {2023},
NUMBER = {6},
ARTICLE-NUMBER = {888},
URL = {https://www.mdpi.com/1099-4300/25/6/888},
PubMedID = {37372232},
ISSN = {1099-4300},
DOI = {10.3390/e25060888}
}

@misc{Yang2024ArXiv,
      title={Robustness, Security, Privacy, Explainability, Efficiency, and Usability of Large Language Models for Code}, 
      author={Zhou Yang and Zhensu Sun and Terry Zhuo Yue and Premkumar Devanbu and David Lo},
      year={2024},
      eprint={2403.07506},
      archivePrefix={arXiv},
      primaryClass={cs.SE}
}

@inproceedings{Perry2023ACM_CCS,
author = {Perry, Neil and Srivastava, Megha and Kumar, Deepak and Boneh, Dan},
title = {Do Users Write More Insecure Code with AI Assistants?},
year = {2023},
isbn = {9798400700507},
publisher = {Association for Computing Machinery},
address = {New York, NY, USA},
url = {https://doi.org/10.1145/3576915.3623157},
doi = {10.1145/3576915.3623157},
booktitle = {Proceedings of the 2023 ACM SIGSAC Conference on Computer and Communications Security},
pages = {2785–2799},
numpages = {15},
location = {Copenhagen, Denmark},
series = {CCS '23}
}

@article{Camara2023SSM,
  author    = {Javier Cámara and Javier Troya and Luis Burgueño and others},
  title     = {On the assessment of generative AI in modeling tasks: an experience report with ChatGPT and UML},
  journal   = {Software and Systems Modeling},
  year      = {2023},
  volume    = {22},
  number    = {3},
  pages     = {781--793},
  doi       = {10.1007/s10270-023-01105-5},
  url       = {https://doi.org/10.1007/s10270-023-01105-5}
}

@inproceedings{Nam2024ACM_ICSE,
author = {Nam, Daye and Macvean, Andrew and Hellendoorn, Vincent and Vasilescu, Bogdan and Myers, Brad},
title = {Using an LLM to Help With Code Understanding},
year = {2024},
isbn = {9798400702174},
publisher = {Association for Computing Machinery},
address = {New York, NY, USA},
url = {https://doi.org/10.1145/3597503.3639187},
doi = {10.1145/3597503.3639187},
booktitle = {Proceedings of the IEEE/ACM 46th International Conference on Software Engineering},
articleno = {97},
numpages = {13},
location = {Lisbon, Portugal},
series = {ICSE '24}
}

@inproceedings{Gu2024ACM_CHI,
author = {Gu, Ken and Grunde-McLaughlin, Madeleine and McNutt, Andrew and Heer, Jeffrey and Althoff, Tim},
title = {How Do Data Analysts Respond to AI Assistance? A Wizard-of-Oz Study},
year = {2024},
isbn = {9798400703300},
publisher = {Association for Computing Machinery},
address = {New York, NY, USA},
url = {https://doi.org/10.1145/3613904.3641891},
doi = {10.1145/3613904.3641891},
booktitle = {Proceedings of the CHI Conference on Human Factors in Computing Systems},
articleno = {1015},
numpages = {22},
location = {Honolulu, HI, USA},
series = {CHI '24}
}

@inproceedings{Nguyen2024ACM_CHI,
author = {Nguyen, Sydney and Babe, Hannah McLean and Zi, Yangtian and Guha, Arjun and Anderson, Carolyn Jane and Feldman, Molly Q},
title = {How Beginning Programmers and Code LLMs (Mis)read Each Other},
year = {2024},
isbn = {9798400703300},
publisher = {Association for Computing Machinery},
address = {New York, NY, USA},
url = {https://doi.org/10.1145/3613904.3642706},
doi = {10.1145/3613904.3642706},
booktitle = {Proceedings of the CHI Conference on Human Factors in Computing Systems},
articleno = {651},
numpages = {26},
location = {Honolulu, HI, USA},
series = {CHI '24}
}

@inproceedings{Rao2024ACM_IUI,
author = {Rao, Nikitha and Tsay, Jason and Kate, Kiran and Hellendoorn, Vincent and Hirzel, Martin},
title = {AI for Low-Code for AI},
year = {2024},
isbn = {9798400705083},
publisher = {Association for Computing Machinery},
address = {New York, NY, USA},
url = {https://doi.org/10.1145/3640543.3645203},
doi = {10.1145/3640543.3645203},
booktitle = {Proceedings of the 29th International Conference on Intelligent User Interfaces},
pages = {837–852},
numpages = {16},
location = {Greenville, SC, USA},
series = {IUI '24}
}

@inproceedings{Jury2024ACM_ACE,
author = {Jury, Breanna and Lorusso, Angela and Leinonen, Juho and Denny, Paul and Luxton-Reilly, Andrew},
title = {Evaluating LLM-generated Worked Examples in an Introductory Programming Course},
year = {2024},
isbn = {9798400716195},
publisher = {Association for Computing Machinery},
address = {New York, NY, USA},
url = {https://doi.org/10.1145/3636243.3636252},
doi = {10.1145/3636243.3636252},
booktitle = {Proceedings of the 26th Australasian Computing Education Conference},
pages = {77–86},
numpages = {10},
location = {Sydney, NSW, Australia},
series = {ACE '24}
}

@misc{Nejjar2024ICSSP,
      title={LLMs for Science: Usage for Code Generation and Data Analysis}, 
      author={Mohamed Nejjar and Luca Zacharias and Fabian Stiehle and Ingo Weber},
      year={2024},
      eprint={2311.16733},
      archivePrefix={arXiv},
      primaryClass={cs.SE},
      url={https://arxiv.org/abs/2311.16733}, 
}

@misc{Liu2023ArXiv,
      title={Which is a better programming assistant? A comparative study between chatgpt and stack overflow}, 
      author={Jinrun Liu and Xinyu Tang and Linlin Li and Panpan Chen and Yepang Liu},
      year={2023},
      eprint={2308.13851},
      archivePrefix={arXiv},
      primaryClass={cs.SE},
      url={https://arxiv.org/abs/2308.13851}, 
}

@inproceedings{Qian2024ACM_IUI,
author = {Qian, Crystal and Wexler, James},
title = {Take It, Leave It, or Fix It: Measuring Productivity and Trust in Human-AI Collaboration},
year = {2024},
isbn = {9798400705083},
publisher = {Association for Computing Machinery},
address = {New York, NY, USA},
url = {https://doi.org/10.1145/3640543.3645198},
doi = {10.1145/3640543.3645198},
booktitle = {Proceedings of the 29th International Conference on Intelligent User Interfaces},
pages = {370–384},
numpages = {15},
location = {Greenville, SC, USA},
series = {IUI '24}
}

@misc{Chopra2023arXiv,
      title={Conversational Challenges in AI-Powered Data Science: Obstacles, Needs, and Design Opportunities}, 
      author={Bhavya Chopra and Ananya Singha and Anna Fariha and Sumit Gulwani and Chris Parnin and Ashish Tiwari and Austin Z. Henley},
      year={2023},
      eprint={2310.16164},
      archivePrefix={arXiv},
      primaryClass={cs.HC},
      url={https://arxiv.org/abs/2310.16164}, 
}

@misc{Erhabor2023arXiv,
      title={Measuring the Runtime Performance of Code Produced with GitHub Copilot}, 
      author={Daniel Erhabor and Sreeharsha Udayashankar and Meiyappan Nagappan and Samer Al-Kiswany},
      year={2023},
      eprint={2305.06439},
      archivePrefix={arXiv},
      primaryClass={cs.SE},
      url={https://arxiv.org/abs/2305.06439}, 
}

@misc{Tan2024arXiv,
      title={How far are AI-powered programming assistants from meeting developers' needs?}, 
      author={Xin Tan and Xiao Long and Xianjun Ni and Yinghao Zhu and Jing Jiang and Li Zhang},
      year={2024},
      eprint={2404.12000},
      archivePrefix={arXiv},
      primaryClass={cs.SE},
      url={https://arxiv.org/abs/2404.12000}, 
}

@misc{Oh2023IEEE_SP,
      title={Poisoned ChatGPT Finds Work for Idle Hands: Exploring Developers' Coding Practices with Insecure Suggestions from Poisoned AI Models}, 
      author={Sanghak Oh and Kiho Lee and Seonhye Park and Doowon Kim and Hyoungshick Kim},
      year={2023},
      eprint={2312.06227},
      archivePrefix={arXiv},
      primaryClass={cs.CR},
      url={https://arxiv.org/abs/2312.06227}, 
}

@misc{Tseng2024arXiv,
      title={Keyframer: Empowering Animation Design using Large Language Models}, 
      author={Tiffany Tseng and Ruijia Cheng and Jeffrey Nichols},
      year={2024},
      eprint={2402.06071},
      archivePrefix={arXiv},
      primaryClass={cs.HC},
      url={https://arxiv.org/abs/2402.06071}, 
}

@article{Sanger2024arXiv,
   title={A qualitative assessment of using ChatGPT as large language model for scientific workflow development},
   volume={13},
   ISSN={2047-217X},
   url={http://dx.doi.org/10.1093/gigascience/giae030},
   DOI={10.1093/gigascience/giae030},
   journal={GigaScience},
   publisher={Oxford University Press (OUP)},
   author={Sänger, Mario and De Mecquenem, Ninon and Lewińska, Katarzyna Ewa and Bountris, Vasilis and Lehmann, Fabian and Leser, Ulf and Kosch, Thomas},
   year={2024} 
}

@misc{kim2024arXiv,
      title={ChatGPT in Data Visualization Education: A Student Perspective}, 
      author={Nam Wook Kim and Hyung-Kwon Ko and Grace Myers and Benjamin Bach},
      year={2024},
      eprint={2405.00748},
      archivePrefix={arXiv},
      primaryClass={cs.HC},
      url={https://arxiv.org/abs/2405.00748}, 
}

@misc{Wang2024arXiv,
      title={Rocks Coding, Not Development--A Human-Centric, Experimental Evaluation of LLM-Supported SE Tasks}, 
      author={Wei Wang and Huilong Ning and Gaowei Zhang and Libo Liu and Yi Wang},
      year={2024},
      eprint={2402.05650},
      archivePrefix={arXiv},
      primaryClass={cs.SE},
      url={https://arxiv.org/abs/2402.05650}, 
}

@misc{Khojah2024arXiv,
      title={Beyond Code Generation: An Observational Study of ChatGPT Usage in Software Engineering Practice}, 
      author={Ranim Khojah and Mazen Mohamad and Philipp Leitner and Francisco Gomes de Oliveira Neto},
      year={2024},
      eprint={2404.14901},
      archivePrefix={arXiv},
      primaryClass={cs.SE},
      url={https://arxiv.org/abs/2404.14901}, 
}

@INPROCEEDINGS{Vasiliniuc2023IEEE_ICCP,
  author={Vasiliniuc, Mircea-Serban and Groza, Adrian},
  booktitle={2023 IEEE 19th International Conference on Intelligent Computer Communication and Processing (ICCP)}, 
  title={Case study: using AI-assisted code generation in mobile teams}, 
  year={2023},
  volume={},
  number={},
  pages={339-346},
  doi={10.1109/ICCP60212.2023.10398656}
}

@inproceedings{Asare2024ACM_ICSE,
author = {Asare, Owura and Nagappan, Meiyappan and Asokan, N.},
title = {A User-centered Security Evaluation of Copilot},
year = {2024},
isbn = {9798400702174},
publisher = {Association for Computing Machinery},
address = {New York, NY, USA},
url = {https://doi.org/10.1145/3597503.3639154},
doi = {10.1145/3597503.3639154},
booktitle = {Proceedings of the IEEE/ACM 46th International Conference on Software Engineering},
articleno = {158},
numpages = {11},
location = {Lisbon, Portugal},
series = {ICSE '24}
}

@misc{Liang2024arXiv,
      title={How LLMs Aid in UML Modeling: An Exploratory Study with Novice Analysts}, 
      author={Beian Wang and Chong Wang and Peng Liang and Bing Li and Cheng Zeng},
      year={2024},
      eprint={2404.17739},
      archivePrefix={arXiv},
      primaryClass={cs.SE},
      url={https://arxiv.org/abs/2404.17739}, 
}

@misc{Chopra2023arXiv1,
      title={Exploring Interaction Patterns for Debugging: Enhancing Conversational Capabilities of AI-assistants}, 
      author={Bhavya Chopra and Yasharth Bajpai and Param Biyani and Gustavo Soares and Arjun Radhakrishna and Chris Parnin and Sumit Gulwani},
      year={2024},
      eprint={2402.06229},
      archivePrefix={arXiv},
      primaryClass={cs.HC},
      url={https://arxiv.org/abs/2402.06229}, 
}

@inproceedings{Liffiton2024ACM_Koli,
author = {Liffiton, Mark and Sheese, Brad E and Savelka, Jaromir and Denny, Paul},
title = {CodeHelp: Using Large Language Models with Guardrails for Scalable Support in Programming Classes},
year = {2024},
isbn = {9798400716539},
publisher = {Association for Computing Machinery},
address = {New York, NY, USA},
url = {https://doi.org/10.1145/3631802.3631830},
doi = {10.1145/3631802.3631830},
booktitle = {Proceedings of the 23rd Koli Calling International Conference on Computing Education Research},
articleno = {8},
numpages = {11},
location = {Koli, Finland},
series = {Koli Calling '23}
}

@inproceedings{Denny2024ACM_SIGCSE,
author = {Denny, Paul and Leinonen, Juho and Prather, James and Luxton-Reilly, Andrew and Amarouche, Thezyrie and Becker, Brett A. and Reeves, Brent N.},
title = {Prompt Problems: A New Programming Exercise for the Generative AI Era},
year = {2024},
isbn = {9798400704239},
publisher = {Association for Computing Machinery},
address = {New York, NY, USA},
url = {https://doi.org/10.1145/3626252.3630909},
doi = {10.1145/3626252.3630909},
booktitle = {Proceedings of the 55th ACM Technical Symposium on Computer Science Education V. 1},
pages = {296–302},
numpages = {7},
location = {Portland, OR, USA},
series = {SIGCSE 2024}
}

@inproceedings{Kazemitabaar2024ACM_CHI,
author = {Kazemitabaar, Majeed and Ye, Runlong and Wang, Xiaoning and Henley, Austin Zachary and Denny, Paul and Craig, Michelle and Grossman, Tovi},
title = {CodeAid: Evaluating a Classroom Deployment of an LLM-based Programming Assistant that Balances Student and Educator Needs},
year = {2024},
isbn = {9798400703300},
publisher = {Association for Computing Machinery},
address = {New York, NY, USA},
url = {https://doi.org/10.1145/3613904.3642773},
doi = {10.1145/3613904.3642773},
booktitle = {Proceedings of the CHI Conference on Human Factors in Computing Systems},
articleno = {650},
numpages = {20},
location = {Honolulu, HI, USA},
series = {CHI '24}
}

@inproceedings{Sheese2024ACM_ACE,
author = {Sheese, Brad and Liffiton, Mark and Savelka, Jaromir and Denny, Paul},
title = {Patterns of Student Help-Seeking When Using a Large Language Model-Powered Programming Assistant},
year = {2024},
isbn = {9798400716195},
publisher = {Association for Computing Machinery},
address = {New York, NY, USA},
url = {https://doi.org/10.1145/3636243.3636249},
doi = {10.1145/3636243.3636249},
booktitle = {Proceedings of the 26th Australasian Computing Education Conference},
pages = {49–57},
numpages = {9},
location = {Sydney, NSW, Australia},
series = {ACE '24}
}

@Article{Kosar2024mdpi,
AUTHOR = {Kosar, Tomaž and Ostojić, Dragana and Liu, Yu David and Mernik, Marjan},
TITLE = {Computer Science Education in ChatGPT Era: Experiences from an Experiment in a Programming Course for Novice Programmers},
JOURNAL = {Mathematics},
VOLUME = {12},
YEAR = {2024},
NUMBER = {5},
ARTICLE-NUMBER = {629},
URL = {https://www.mdpi.com/2227-7390/12/5/629},
ISSN = {2227-7390},
DOI = {10.3390/math12050629}
}

@inproceedings{Choudhuri2024ACM_ICSE,
author = {Choudhuri, Rudrajit and Liu, Dylan and Steinmacher, Igor and Gerosa, Marco and Sarma, Anita},
title = {How Far Are We? The Triumphs and Trials of Generative AI in Learning Software Engineering},
year = {2024},
isbn = {9798400702174},
publisher = {Association for Computing Machinery},
address = {New York, NY, USA},
url = {https://doi.org/10.1145/3597503.3639201},
doi = {10.1145/3597503.3639201},
booktitle = {Proceedings of the IEEE/ACM 46th International Conference on Software Engineering},
articleno = {184},
numpages = {13},
location = {Lisbon, Portugal},
series = {ICSE '24}
}

@inproceedings{Xiao2024ACM_CHI,
author = {Xiao, Ruiwei and Hou, Xinying and Stamper, John},
title = {Exploring How Multiple Levels of GPT-Generated Programming Hints Support or Disappoint Novices},
year = {2024},
isbn = {9798400703317},
publisher = {Association for Computing Machinery},
address = {New York, NY, USA},
url = {https://doi.org/10.1145/3613905.3650937},
doi = {10.1145/3613905.3650937},
booktitle = {Extended Abstracts of the 2024 CHI Conference on Human Factors in Computing Systems},
articleno = {142},
numpages = {10},
location = {
},
series = {CHI EA '24}
}

@inproceedings{Kuramitsu2023ACM_SPLASH-E,
author = {Kuramitsu, Kimio and Obara, Yui and Sato, Miyu and Obara, Momoka},
title = {KOGI: A Seamless Integration of ChatGPT into Jupyter Environments for Programming Education},
year = {2023},
isbn = {9798400703904},
publisher = {Association for Computing Machinery},
address = {New York, NY, USA},
url = {https://doi.org/10.1145/3622780.3623648},
doi = {10.1145/3622780.3623648},
booktitle = {Proceedings of the 2023 ACM SIGPLAN International Symposium on SPLASH-E},
pages = {50–59},
numpages = {10},
location = {Cascais, Portugal},
series = {SPLASH-E 2023}
}

@inproceedings{Xue2024ACM_ICSE,
author = {Xue, Yuankai and Chen, Hanlin and Bai, Gina R. and Tairas, Robert and Huang, Yu},
title = {Does ChatGPT Help With Introductory Programming?An Experiment of Students Using ChatGPT in CS1},
year = {2024},
isbn = {9798400704987},
publisher = {Association for Computing Machinery},
address = {New York, NY, USA},
url = {https://doi.org/10.1145/3639474.3640076},
doi = {10.1145/3639474.3640076},
booktitle = {Proceedings of the 46th International Conference on Software Engineering: Software Engineering Education and Training},
pages = {331–341},
numpages = {11},
location = {Lisbon, Portugal},
series = {ICSE-SEET '24}
}

@inproceedings{Arawjo2024ACM_CHI,
author = {Arawjo, Ian and Swoopes, Chelse and Vaithilingam, Priyan and Wattenberg, Martin and Glassman, Elena L.},
title = {ChainForge: A Visual Toolkit for Prompt Engineering and LLM Hypothesis Testing},
year = {2024},
isbn = {9798400703300},
publisher = {Association for Computing Machinery},
address = {New York, NY, USA},
url = {https://doi.org/10.1145/3613904.3642016},
doi = {10.1145/3613904.3642016},
booktitle = {Proceedings of the CHI Conference on Human Factors in Computing Systems},
articleno = {304},
numpages = {18},
location = {Honolulu, HI, USA},
series = {CHI '24}
}

@inproceedings{Vaithilingam2024ACM_CHI,
author = {Vaithilingam, Priyan and Glassman, Elena L. and Inala, Jeevana Priya and Wang, Chenglong},
title = {DynaVis: Dynamically Synthesized UI Widgets for Visualization Editing},
year = {2024},
isbn = {9798400703300},
publisher = {Association for Computing Machinery},
address = {New York, NY, USA},
url = {https://doi.org/10.1145/3613904.3642639},
doi = {10.1145/3613904.3642639},
booktitle = {Proceedings of the CHI Conference on Human Factors in Computing Systems},
articleno = {985},
numpages = {17},
location = {Honolulu, HI, USA},
series = {CHI '24}
}

@inproceedings{Yan2024ACM_CHI,
author = {Yan, Litao and Hwang, Alyssa and Wu, Zhiyuan and Head, Andrew},
title = {Ivie: Lightweight Anchored Explanations of Just-Generated Code},
year = {2024},
isbn = {9798400703300},
publisher = {Association for Computing Machinery},
address = {New York, NY, USA},
url = {https://doi.org/10.1145/3613904.3642239},
doi = {10.1145/3613904.3642239},
booktitle = {Proceedings of the CHI Conference on Human Factors in Computing Systems},
articleno = {140},
numpages = {15},
location = {Honolulu, HI, USA},
series = {CHI '24}
}

@ARTICLE{Wang2024IEEE_TVCG,
  author={Wang, Chenglong and Thompson, John and Lee, Bongshin},
  journal={IEEE Transactions on Visualization and Computer Graphics}, 
  title={Data Formulator: AI-Powered Concept-Driven Visualization Authoring}, 
  year={2024},
  volume={30},
  number={1},
  pages={1128-1138},
  doi={10.1109/TVCG.2023.3326585}
}

@inproceedings{Dirin2023rG,
author = {Dirin, Amir and Laine, Teemu},
year = {2024},
month = {05},
pages = {},
booktitle = {CSEDU24},
title = {Examining the Utilization of Artificial Intelligence Tools by Students in Software Engineering Projects},
doi = {10.5220/0012729400003693}
}

@misc{Fakhoury2024arXiv,
      title={LLM-based Test-driven Interactive Code Generation: User Study and Empirical Evaluation}, 
      author={Sarah Fakhoury and Aaditya Naik and Georgios Sakkas and Saikat Chakraborty and Shuvendu K. Lahiri},
      year={2024},
      eprint={2404.10100},
      archivePrefix={arXiv},
      primaryClass={cs.SE},
      url={https://arxiv.org/abs/2404.10100}, 
}

@misc{Chatterjee2024SEC,
      title={The Impact of AI Tool on Engineering at ANZ Bank An Empirical Study on GitHub Copilot within Corporate Environment}, 
      author={Sayan Chatterjee and Ching Louis Liu and Gareth Rowland and Tim Hogarth},
      year={2024},
      eprint={2402.05636},
      archivePrefix={arXiv},
      primaryClass={cs.SE},
      url={https://arxiv.org/abs/2402.05636}, 
}

@ARTICLE{Liu2024IEEE_TSE,
  author={Liu, Jiaqi and Zhang, Fengming and Zhang, Xin and Yu, Zhiwen and Wang, Liang and Zhang, Yao and Guo, Bin},
  journal={IEEE Transactions on Software Engineering}, 
  title={hmCodeTrans: Human–Machine Interactive Code Translation}, 
  year={2024},
  volume={50},
  number={5},
  pages={1163-1181},
  doi={10.1109/TSE.2024.3379583}
}

@misc{Yin2024arXiv,
      title={Exploring the Potential of Large Language Models in Artistic Creation: Collaboration and Reflection on Creative Programming}, 
      author={Anqi Wang and Zhizhuo Yin and Yulu Hu and Yuanyuan Mao and Pan Hui},
      year={2024},
      eprint={2402.09750},
      archivePrefix={arXiv},
      primaryClass={cs.HC},
      url={https://arxiv.org/abs/2402.09750}, 
}

@inproceedings{Feng2024ACM_ICSE,
author = {Feng, Sidong and Chen, Chunyang},
title = {Prompting Is All You Need: Automated Android Bug Replay with Large Language Models},
year = {2024},
isbn = {9798400702174},
publisher = {Association for Computing Machinery},
address = {New York, NY, USA},
url = {https://doi-org.proxy.lib.uwaterloo.ca/10.1145/3597503.3608137},
doi = {10.1145/3597503.3608137},
booktitle = {Proceedings of the IEEE/ACM 46th International Conference on Software Engineering},
articleno = {67},
numpages = {13},
location = {Lisbon, Portugal},
series = {ICSE '24}
}

@inproceedings{Su2024ACM_ICSE,
author = {Su, Yanqi and Liao, Dianshu and Xing, Zhenchang and Huang, Qing and Xie, Mulong and Lu, Qinghua and Xu, Xiwei},
title = {Enhancing Exploratory Testing by Large Language Model and Knowledge Graph},
year = {2024},
isbn = {9798400702174},
publisher = {Association for Computing Machinery},
address = {New York, NY, USA},
url = {https://doi-org.proxy.lib.uwaterloo.ca/10.1145/3597503.3639157},
doi = {10.1145/3597503.3639157},
booktitle = {Proceedings of the IEEE/ACM 46th International Conference on Software Engineering},
articleno = {98},
numpages = {12},
location = {Lisbon, Portugal},
series = {ICSE '24}
}

@misc{Patton2024arXiv,
      title={Aptly: Making Mobile Apps from Natural Language}, 
      author={Evan W. Patton and David Y. J. Kim and Ashley Granquist and Robin Liu and Arianna Scott and Jennet Zamanova and Harold Abelson},
      year={2024},
      eprint={2405.00229},
      archivePrefix={arXiv},
      primaryClass={cs.HC},
      url={https://arxiv.org/abs/2405.00229}, 
}

@inproceedings{Karli2024ACM_HRI,
author = {Karli, Ulas Berk and Chen, Juo-Tung and Antony, Victor Nikhil and Huang, Chien-Ming},
title = {Alchemist: LLM-Aided End-User Development of Robot Applications},
year = {2024},
isbn = {9798400703225},
publisher = {Association for Computing Machinery},
address = {New York, NY, USA},
url = {https://doi-org.proxy.lib.uwaterloo.ca/10.1145/3610977.3634969},
doi = {10.1145/3610977.3634969},
booktitle = {Proceedings of the 2024 ACM/IEEE International Conference on Human-Robot Interaction},
pages = {361–370},
numpages = {10},
location = {Boulder, CO, USA},
series = {HRI '24}
}

@misc{Denny2024iTCSE,
      title={Explaining Code with a Purpose: An Integrated Approach for Developing Code Comprehension and Prompting Skills}, 
      author={Paul Denny and David H. Smith IV au2 and Max Fowler and James Prather and Brett A. Becker and Juho Leinonen},
      year={2024},
      eprint={2403.06050},
      archivePrefix={arXiv},
      primaryClass={cs.HC},
      url={https://arxiv.org/abs/2403.06050}, 
}

@misc{Arora2024arXiv,
      title={Analyzing LLM Usage in an Advanced Computing Class in India}, 
      author={Chaitanya Arora and Utkarsh Venaik and Pavit Singh and Sahil Goyal and Jatin Tyagi and Shyama Goel and Ujjwal Singhal and Dhruv Kumar},
      year={2024},
      eprint={2404.04603},
      archivePrefix={arXiv},
      primaryClass={cs.HC},
      url={https://arxiv.org/abs/2404.04603}, 
}

@misc{Yan2024arXiv,
      title={IntelliExplain: Enhancing Interactive Code Generation through Natural Language Explanations for Non-Professional Programmers}, 
      author={Hao Yan and Thomas D. Latoza and Ziyu Yao},
      year={2024},
      eprint={2405.10250},
      archivePrefix={arXiv},
      primaryClass={cs.HC},
      url={https://arxiv.org/abs/2405.10250}, 
}

@INPROCEEDINGS{Ouaazki2023TALE,
  author={Ouaazki, Abdessalam and Bergram, Kristoffer and Holzer, Adrian},
  booktitle={2023 IEEE International Conference on Teaching, Assessment and Learning for Engineering (TALE)}, 
  title={Leveraging ChatGPT to Enhance Computational Thinking Learning Experiences}, 
  year={2023},
  volume={},
  number={},
  pages={1-7},
  doi={10.1109/TALE56641.2023.10398358}
}

@article{Sun2024IJETHE,
  author    = {Dongfang Sun and Azzeddine Boudouaia and Cheng Zhu and others},
  title     = {Would ChatGPT-facilitated programming mode impact college students’ programming behaviors, performances, and perceptions? An empirical study},
  journal   = {International Journal of Educational Technology in Higher Education},
  year      = {2024},
  volume    = {21},
  number    = {1},
  pages     = {14},
  doi       = {10.1186/s41239-024-00446-5},
  url       = {https://doi.org/10.1186/s41239-024-00446-5}
}

@article{Jing2024HSSC,
  author    = {Yong Jing and Hao Wang and Xinyu Chen and others},
  title     = {What factors will affect the effectiveness of using ChatGPT to solve programming problems? A quasi-experimental study},
  journal   = {Humanities and Social Sciences Communications},
  year      = {2024},
  volume    = {11},
  number    = {1},
  pages     = {319},
  doi       = {10.1057/s41599-024-02751-w},
  url       = {https://doi.org/10.1057/s41599-024-02751-w}
}

@article{Johnson2024JRTC,
  author    = {Daniel M. Johnson and William Doss and Christopher M. Estepp},
  title     = {Using ChatGPT with Novice Arduino Programmers: Effects on Performance, Interest, Self-Efficacy, and Programming Ability},
  journal   = {Journal of Research in Technical Careers},
  year      = {2024},
  volume    = {8},
  number    = {1},
  doi       = {10.9741/2578-2118.1152},
  url       = {https://doi.org/10.9741/2578-2118.1152}
}

@misc{Bernstein2024ITiCSE,
      title={"Like a Nesting Doll": Analyzing Recursion Analogies Generated by CS Students using Large Language Models}, 
      author={Seth Bernstein and Paul Denny and Juho Leinonen and Lauren Kan and Arto Hellas and Matt Littlefield and Sami Sarsa and Stephen MacNeil},
      year={2024},
      eprint={2403.09409},
      archivePrefix={arXiv},
      primaryClass={cs.HC},
      url={https://arxiv.org/abs/2403.09409}, 
}

@INPROCEEDINGS{Aillon2023IEEE_C3,
  author={Aillon, Santiago and Garcia, Alejandro and Velandia, Nicolas and Zarate, Daniel and Wightman, Pedro},
  booktitle={2023 IEEE Colombian Caribbean Conference (C3)}, 
  title={Empirical evaluation of automated code generation for mobile applications by AI tools}, 
  year={2023},
  volume={},
  number={},
  pages={1-6},
  doi={10.1109/C358072.2023.10436306}
}

@article{Korkmaz2017CHB,
title = {A validity and reliability study of the computational thinking scales (CTS)},
journal = {Computers in Human Behavior},
volume = {72},
pages = {558-569},
year = {2017},
issn = {0747-5632},
doi = {https://doi.org/10.1016/j.chb.2017.01.005},
url = {https://www.sciencedirect.com/science/article/pii/S0747563217300055},
author = {Özgen Korkmaz and Recep Çakir and M. Yaşar Özden}
}

@standard{IEEE730,
  title = {IEEE Standard for Software Quality Assurance Processes},
  organization = {IEEE},
  year = {2014},
  url = {https://standards.ieee.org/standard/730-2014.html}
}

@standard{ieee1061,
  title = {IEEE Standard for a Software Quality Metrics Methodology},
  organization = {IEEE},
  year = {1998},
  url = {https://standards.ieee.org/standard/1061-1998.html}
}

@misc{cert_secure_coding,
  title = {CERT Secure Coding Standards},
  author = {{Computer Emergency Response Team}},
  url = {https://www.securecoding.cert.org/}
}

@misc{white2023arXiv,
      title={A Prompt Pattern Catalog to Enhance Prompt Engineering with ChatGPT}, 
      author={Jules White and Quchen Fu and Sam Hays and Michael Sandborn and Carlos Olea and Henry Gilbert and Ashraf Elnashar and Jesse Spencer-Smith and Douglas C. Schmidt},
      year={2023},
      eprint={2302.11382},
      archivePrefix={arXiv},
      primaryClass={cs.SE},
      url={https://arxiv.org/abs/2302.11382}, 
}

\end{document}